\documentclass[aps,prb,reprint,groupedaddress]{revtex4-2}
\usepackage{graphicx}
\usepackage{dcolumn}
\usepackage{bm}

\usepackage[utf8]{inputenc}
\usepackage[T1]{fontenc}
\usepackage{booktabs, array, mathptmx, float, tabularx, booktabs, lipsum, amsmath,multirow}
\usepackage{siunitx, xcolor}
\usepackage[version=4]{mhchem}
\graphicspath{{figs/}{figsgaoerb/}} 
\usepackage[colorlinks,linkcolor=blue,anchorcolor=blue,citecolor=blue]{hyperref}

\begin{document}


\title{Doping-driven evolution of pairing symmetry in pressurized La$_3$Ni$_2$O$_7$}


\author{Hai-Yang Zhang}
\email[]{haiyangzhang03@163.com}
\author{Yu-Jie Bai}
\author{Fan-Jie Kong}
\affiliation{Department of Physics, Yancheng Institute of Technology, Yancheng 224051, China}


\date{\today}

\begin{abstract}
	We investigate the superconducting pairing symmetry and its doping evolution in pressurized  La$_3$Ni$_2$O$_7$. For the undoped compound, the most favorable pairing state is found to be $s_{\pm}$-wave, characterized by  sign reversal of the gap functions between the Fermi pockets. A detailed analysis of the pairing interactions reveals that this unconventional state originates from repulsive interactions mediated by the magnetic odd modes of the bilayer nickelate. Upon hole doping, the $\gamma$ Fermi pocket expands, which amplifies the intrapocket repulsions on this pocket. These repulsions, driven by the magnetic even modes, gradually dominates the pairing interactions and ultimately drive a transition in pairing symmetry from $s_{\pm}$-wave to $d_{xy}$-wave in the heavily hole-doped regime. In stark contrast, the $s_{\pm}$-wave pairing persists under electron doping, even deep into the heavily electron-doped regime where a Lifshitz transition occurs. This finding suggests that the $\gamma$ Fermi pocket is not essential for the emergence of superconductivity in bilayer nickelates. In fact, the $s_{\pm}$-wave pairing becomes more robust in the absence of the $\gamma$ pocket, as spin fluctuations are strongly enhanced by the favorable nesting between the $\alpha$ and $\beta$ pockets --- a condition guaranteed by Luttinger's theorem and the Fermi surface topology. We believe that exploring the doping evolution of superconducting pairing will open a new realm for testing the pairing mechanism in pressurized La$_3$Ni$_2$O$_7$.    
\end{abstract}


\maketitle

\section{The introduction}
The discovering of superconductivity above $80$ K in pressurized La$_3$Ni$_2$O$_7$ has ignited intense interest in condensed matter community~\cite{SunNat}. Soon after, both zero resistance and the Meissner effect were observed in single-crystalline~\cite{zerofilamentary,zeroR1,zeroR2,zeroR4,zeroRandMesn} and polycrystalline~\cite{zeroRpolycrystal} samples under hydrostatic pressure. Study also shows that  superconductivity can remain stable up to $90$ GPa, exhibiting a right-triangle-like shape in the pressure-temperature phase diagram~\cite{zeroR4}. In recent work, the high-quality single crystals have been synthesized, and their superconducting transition temperature under high pressure exceeds $90$K~\cite{bulk90K}. Similar to cuprates~\cite{CupratesShenZhiXunRMP,CupratesZongshu} and the iron-based superconductors~\cite{ScalapinoRMP}, experiments have revealed that superconductivity is in close proximity to a spin-density-wave (SDW) phase, which occurs around $150$ K in the parent compounds~\cite{FengRIXS,uSR,uSRsplitDW,ChenxianhuiNMR}. Density functional theory (DFT) calculations indicate that the Ni-$t_{2g}$ orbitals are fully occupied, and the low-energy bands are dominated by the Ni-$3d_{z^{2}}$ and Ni-$3d_{x^2-y^2}$ orbitals. It has been suggested that the applied pressure lifts the Ni-$3d_{z^{2}}$ band across the Fermi level, forming a hole Fermi pocket that is believed to be essential for the emergence of superconductivity~\cite{SunNat,zeroR1}. The DFT band structure and the resulting Fermi surfaces are consistent with ARPES measurements on samples at ambient pressure~\cite{ZXJNCarpes,ARPEScpl}. To date, extensive studies have been carried out to investigate the superconducting mechanism of pressurized bilayer nickelates.  Generally,  theoretical proposals fall into two categories. One class originates from the strong coupling limit, where the localized Ni-$3d_{z^{2}}$ orbitals are antiferromagnetically correlated between the upper and lower layers. Superconductivity emerges via Hund's coupling or hybrization with the itinerant Ni-$3d_{x^{2}-y^{2}}$ electrons ~\cite{tJDMRGzhanggm,tJKunJiang,tJLange,tJYang,tJzhangfchun,tJzhangyh,tJXiangTaoDwave,LuChenWuCprl,BilayerSu,Wucongcpl2024,Yangyifeng,YaodaotJ,WangFaCPLpairing}. Depending on the specific model and parameters, the predicted pairing symmetry ranges from the $s$-wave and $d$-wave to even $d+is$-wave. The other class is based on weak coupling theories~\cite{FLEXsaka,rpaZhangyangSwave,EreminRPAandpairing,YangfanPRLOdefi,rpazhangyangNC,rpaBraz,FRGwang,FRGwu,FRGwu2026,FLEXdvs}, where magnetic fluctuations are proposed to be responsible for the emergence of superconductivity. Most of these theories give rise to $s_{\pm}$-wave pairing. Currently, the role of the $\gamma$ pocket, dominated by the Ni-$3d_{z^{2}}$ orbital, is hotly debated. In some strong coupling theories~\cite{LuChenWuCprl,BilayerSu,Wucongcpl2024,Yangyifeng}, the emergence of the $\gamma$ Fermi pocket is not particularly relevant, and superconductivity is attributed primarily to the itinerant Ni-$3d_{x^2-y^2}$ orbital. Conversely, in other theories, especially those based on weak coupling~\cite{EreminRPAandpairing,YangfanPRLOdefi,rpazhangyangNC,rpaBraz,FRGwang,FRGwu,FRGwu2026,FLEXdvs}, the emergence of the $\gamma$ Fermi pocket is believed to stabilize superconductivity.

As demonstrated in cuprates~\cite{CupratesShenZhiXunRMP,CupratesZongshu} and iron-based superconductors~\cite{ScalapinoRMP}, the pressurized bilayer nickelate is expected to exhibit a rich evolution of quantum states with carrier doping,  thus offering a valuable platform to study the mechanism of superconductivity. Therefore, investigating the effects of carrier doping is highly desirable, as it may reveal the origin of superconductivity and its intricate relationship with the normal-state density-wave orders~\cite{FengRIXS,uSR,uSRsplitDW,ChenxianhuiNMR}. In this paper, we adopt the minimal bilayer-two-orbital model from Ref.~\cite{Yaotbmodel} to study the doping evolution of the pairing symmetry in pressurized La$_3$Ni$_2$O$_7$. Consistent with previous theoretical studies~\cite{YangfanPRLOdefi,rpazhangyangNC,rpaBraz,FRGwang,FRGwu,FRGwu2026}, for a filling of $n=3.0$ with realistic interaction parameters, the most favorable pairing state is the $s_{\pm}$-wave state, characterized by sign reversal of the gap functions between the Fermi pockets. By examining the dependence of the pairing scattering interaction on Fermi momenta, we identify that this $s_{\pm}$-wave pairing state is directly tied to the magnetic odd modes of the bilayer nickelate -- modes that were established in a previous study~\cite{zhangnjp}. There exist competing $d_{xy}$-wave and $d_{x^2-y^2}$-wave pairing states, which are nearly degenerate with the $s_{\pm}$-wave state when the interaction parameter is small. However, as the interaction strength increases, the $s_{\pm}$-wave pairing state becomes  dominant. We further investigate the doping evolution of the pairing state. In the hole-doped scenario, as the $\gamma$ pocket expands, the intrapocket repulsion driven by the magnetic even modes, is strongly enhanced by the intensified particle-hole scatterings within this pocket, gradually dominating the pairing interactions. Consequently, sign reversal of the gap function emerges on the $\gamma$ Fermi surface itself, driving a transition in pairing symmetry from $s_{\pm}$-wave to $d_{xy}$-wave. In contrast, for the electron-doped scenario, we find that the $s_{\pm}$-wave pairing state persists even across the Lifshitz transition point. That is, it survives in the absence of the hole $\gamma$ Fermi pocket, whose role in the superconductivity of bilayer nickelates has been highlighted by previous studies~\cite{SunNat,EreminRPAandpairing,YangfanPRLOdefi,rpazhangyangNC,rpaBraz,FRGwang,FRGwu,FRGwu2026,FLEXdvs}. In fact, below the Lifshitz transition, the spin fluctuations become stronger, as guaranteed by the Luttinger's theorem and the Fermi surface topology~\cite{zhangnjp}. Naturally, if spin fluctuations are considered as the pairing glue, the $s_{\pm}$-wave pairing state becomes more robust. Thus, we emphasize that the $\gamma$ Fermi pocket is not essential for the emergence of superconductivity in the bilayer nickelate, given that the pairing scattering interactions are tightly related to the behaviors of the magnetic excitation modes. These theoretical predictions await experimental verification in future investigations. Recently, bilayer nickelate thin films have been successfully synthesized on substrates~\cite{Nie2026film,Chen2026film,XQK2025film,Hao2025film,Szx2026PRXfilm,NieYF2026NPfilm,Nie2026film}. The strong compressive epitaxial strain, induced by the substrate, mimics the effect of hydrostatic pressure, thereby enabling the study of the physical properties of superconducting nickelates at ambient pressure. Complementary investigations of thin-film and bulk samples therefore provide a powerful route toward elucidating the microscopic pairing mechanism in bilayer nickelate superconductors.

\section{The model and formulas}
The minimal bilayer-two-orbital Hamiltonian $H=H_{0}+H_{I}$ is adopted to carry out the calculations. The tight-binding part reads $H_{0}=\sum_{kab\sigma}^{LJ}\epsilon_{ab}^{LJ}(k)C^{+}_{La\sigma}(k)C_{Jb\sigma}(k)$~\cite{Yaotbmodel}. Here, $L$ and $J$ denote the layer indices, while $a$, $b$ and $\sigma$ refer to the orbital and spin indices, respectively. For brevity,  $x$ and $z$ are used as shorthand for the Ni-$3d_{x^2-y^2}$ and Ni-$3d_{z^2}$ orbitals. With this notation, and following Ref.~\cite{Yaotbmodel}, the kinetic energy terms are given by $\epsilon_{xx/zz}^{LL}(k)=-2t_{1}^{x/z}\gamma_{k}-4t_{2}^{x/z}\gamma_{k}^{'}+
\epsilon^{x/z}-\mu$, $\epsilon_{xz}^{LL}(k)=2t_{3}^{xz}\gamma_{k}^{''}$, $\epsilon_{xx/zz}^{L\bar{L}}(k)=2t_{\perp}^{x/z}$ , $\epsilon_{xz}^{L\bar{L}}(k)=2t_{4}^{xz}\gamma_{k}^{''}$, where $\gamma_{k}=$cos $k_x+$cos $k_y$, $\gamma_{k}^{'}=$cos $k_x$cos $k_y$ and $\gamma_{k}^{''}=$cos $k_x-$cos $k_y$. The hopping integrals, together with the onsite energies, taking the following values (as listed in Ref.~\cite{Yaotbmodel}) : $t_{1}^{x}=-0.483$, $t_{1}^{z}=-0.110$, $t_{2}^{x}=0.069$, $t_{2}^{z}=-0.017$, $t_{3}^{xz}=0.239$, $t_{\perp}^{x}=0.005$, $t_{\perp}^{z}=-0.635$, $t_{4}^{xz}=-0.034$, $\epsilon^{x}=0.776$ and $\epsilon^{z}=0.409$.

The interacting part $H_{I}$ of the Hamiltonian takes the form
\begin{eqnarray}
	H_{I}&=&U\sum_{L,i,a} n_{Lia\uparrow}n_{Lia\downarrow}+U'\sum_{L,i,a<b}n_{Lia}n_{Lib} \nonumber\\
	&+&J\sum_{L,i,a<b}
	C^{+}_{Lia\sigma}C^{+}_{Lib\sigma^{'}}C_{Lia\sigma^{'}}C_{Lib\sigma} \nonumber\\
	&+&J'\sum_{L,i,a\neq b} C^{+}_{Lia\uparrow}C^{+}_{Lia\downarrow}
	C_{Lib\downarrow}C_{Lib\uparrow}.
	\label{eqn.1}
\end{eqnarray}
Here, $n_{Lia}=n_{Lia\uparrow}+n_{Lia\downarrow}$ counts the electrons residing in orbital-$a$ on site $i$ within layer $L$. In this expression, $U$, $U'$, $J$, $J'$ represent the strengths of the intraorbital interaction, interorbital
interaction, Hund coupling and pair hopping terms, respectively.
Throughout this work, the constrains $U=U'+J+J'$ and $J=J'$ are imposed,  as required by rotational symmetry in both spatial and spin space.

Within the framework of the random-phase approximation (RPA), we study the superconducting pairing symmetry of the bilayer nickelate La$_3$Ni$_{2}$O$_{7}$ under high pressure, as well as its evolution with doping. In the present model, the single-particle Green's function is defined as $\hat{G}_{ab\sigma}^{LJ}(k,\tau)=-\langle T[C_{ La\sigma}(k,\tau)C_{Jb\sigma}^{+}(k,0)]\rangle$. The spin and charge susceptibilities are defined, respectively, as
$\hat{\mathcal{X}}_{ab,cd}^{S,LJ,KM}(q,\tau)=\langle T[S^{-}_{La,Jb}(q,\tau)S_{Kc,Md}^{+}(q,0)]\rangle$ and $\hat{\mathcal{X}}_{ab,cd}^{C,LJ,KM}(q,\tau)=\langle T[\rho_{La,Jb}(q,\tau)\rho_{Kc,Md}(q,0)]\rangle$. Here $S^{-}_{La,Jb}(q)=\frac{1}{\sqrt{N}}\sum_{k}C_{La\downarrow}^{+}(k)C_{Jb\uparrow}(k+q)$ and $S^{+}_{La,Jb}(q)=\frac{1}{\sqrt{N}}\sum_{k}C_{La\uparrow}^{+}(k+q)C_{Jb\downarrow}(k)$ are the spin lowering and raising operators, while $\rho_{La,Jb}(q)=\frac{1}{\sqrt{N}}\sum_{k\sigma}C^{+}_{La\sigma}(k+q)C_{Jb\sigma}(k)$ is the charge density operator.
The bare spin and charge susceptibilities can be expressed as  $\hat{\chi}^{LJ,KM}_{ab,cd}(q,\tau)=-\frac{1}{N}\sum_{k}G_{da}^{ML}(k,-\tau)G_{bc}
^{JK}(k+q,\tau)$.
After Fourier transformation, we obtain $\hat{\chi}_{ab,cd}^{LJ,KM}(q,i\omega_{n})=-\frac{1}{N\beta}\sum_{k,i\omega_{m}}G_{da}
^{ML}(k,i\omega_{m})G_{bc}^{JK}(k+q,i\omega_{m}+i\omega_{n})$,
where $\beta=\frac{1}{T}$ is the inverse temperature.

When decomposed into the spin channel, the nonzero entries of the interaction matrix read $\hat{U}_{aa,aa}^{S,LL,LL}(q)=U$, $\hat{U}_{aa,bb(a\neq b)}^{S,LL,LL}(q)=
J$, $\hat{U}_{ab,ba(a\neq b)}^{S,LL,LL}(q)=U'$ and $\hat{U}_{ab,ab(a\neq b)}^{S,LL,LL}(q)=J'$. Similarly, the nonzero entries of the charge interaction matrix read as $\hat{U}_{aa,aa}^{C,LL,LL}(q)=U$, $\hat{U}_{aa,bb(a\neq b)}^{C,LL,LL}(q)=
2U'-J$, $\hat{U}_{ab,ba(a\neq b)}^{C,LL,LL}(q)=2J-U'$ and $\hat{U}_{ab,ab(a\neq b)}^{C,LL,LL}(q)=J'$. In this way, the RPA spin and charge susceptibilities are then given by $\hat{\mathcal{X}}^{S}(q,i\omega_{n})=\hat{\chi}(q,i\omega_{n})\bigl(\hat{I}-\hat{U}^{S}(q)\hat{\chi}
(q,i\omega_{n})\bigr)^{-1}$ and $\hat{\mathcal{X}}^{C}(q,i\omega_{n})=\hat{\chi}(q,i\omega_{n})\bigl(\hat{I}+\hat{U}^{C}(q)\hat{\chi}
(q,i\omega_{n})\bigr)^{-1}$, respectively. Upon analytic continuation, the retarded spin and charge susceptibility $\hat{\mathcal{X}}^{R,S}(q,\omega)=\hat{\mathcal{X}}^{S}(q,\omega+i\Gamma)$ and  $\hat{\mathcal{X}}^{R,C}(q,\omega)=\hat{\mathcal{X}}^{C}(q,\omega+i\Gamma)$ are obtained, where $\Gamma$ is the broadening factor, set to $0.02$ in this paper. The zero-frequency spin and charge susceptibility $\hat{\mathcal{X}}^{R,S}(q)$ and $\hat{\mathcal{X}}^{R,C}(q)$ are then used to construct the pairing interaction in the standard way. 

In this paper, we focus on the spin-singlet pairing of pressurized La$_3$Ni$_2$O$_7$~\cite{tJDMRGzhanggm,tJKunJiang,tJLange,tJYang,tJzhangfchun,tJzhangyh,tJXiangTaoDwave,LuChenWuCprl,BilayerSu,Wucongcpl2024,Yangyifeng,YaodaotJ,WangFaCPLpairing,FLEXsaka,rpaZhangyangSwave,EreminRPAandpairing,YangfanPRLOdefi,rpazhangyangNC,rpaBraz,FRGwang,FRGwu,FRGwu2026,FLEXdvs}. Decomposing these fluctuation-mediated and bare interactions into the singlet pairing channel, the pairing interaction is obtained as $V^{p}_{ij,lm}(k,k')=\frac{1}{2}[ 3\hat{U}^{S}\hat{\mathcal{X}}^{R,S}(k-k')\hat{U}^{S}- \hat{U}^{C}\hat{\mathcal{X}}^{R,C}(k-k')\hat{U}^{C}+\hat{U}^{S}+\hat{U}^{C}]_{im,jl}$, where $i,j,l,m$ are the combined layer and orbital indices. Projecting this matrix onto the band representation yields $V^{p}_{\mu\nu}(k,k')=\sum_{ijlm}\phi^{*}_{\mu}(k,i)\phi^{*}_{\mu}(k,j)\phi_{\nu}(k',l)\phi_{\nu}(k',m)V^{p}_{ij,lm}(k,k')$ with $\mu,\nu$ the band indices and $\phi_{\mu}$, $\phi_{\nu}$ the wave-function of the Bloch electrons. Owing to the singlet pairing nature, the interaction matrix further reads $\Gamma_{\mu\nu}(k,k')=\frac{1}{2}[V^{p}_{\mu\nu}(k,k')+V^{p}_{\mu\nu}(k,-k')]$, where the momenta $k$ and $k'$ are restricted to the Fermi surfaces of the $\alpha$, $\beta$ and $\gamma$ pockets, as before. Diagonalizing this pairing matrix via $\frac{1}{(2\pi)^2}\int_{FS}\frac{dk'}{\lvert v_{F}(k') \rvert}\Gamma(k,k')g_{p}(k')=-\lambda_{p}g_{p}(k)$ yields the pairing strengths $\lambda_{p}$ for various channels and the corresponding eigenfunction $g_{p}(k)$ of the Cooper pairs~\cite{rpaZhangyangSwave,rpazhangyangNC}.  

Throughout the paper, $T$ is set to $0.001$. All energies are in eV. All calculations are performed on a $256\times256\times2$ lattice.

\section{The numerical results}
\subsection{Pairing symmetry for the pressurized La$_3$Ni$_2$O$_7$ with $n=3.0$}
\begin{figure}
	\centering\includegraphics[width=0.45\textwidth]{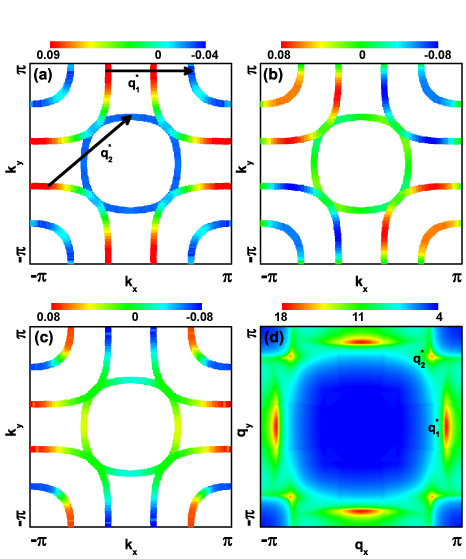}
	\caption{(Color online) 
		(a), (b) and (c) are the first three leading pairing states of $s_{\pm}$, $d_{xy}$ and $d_{x^2-y^2}$ symmetries, respectively. (d) shows the momentum dependence of the static odd-channel spin susceptibility $\hat{\mathcal{X}}^{R,S}(q)$.}
	\label{f1}
\end{figure}

In Fig.~\ref{f1}, we show the three leading pairing channels for $U=1.2$ and $J=U/6$ at the filling $n=3.0$. This $J/U$ ratio is realistic, as indicated by constrained RPA calculation~\cite{CorrelationJU}. The value of $U$ is chosen to place the system near the SDW instability, consistent with the experimental observations~\cite{FengRIXS,uSR,uSRsplitDW,ChenxianhuiNMR}. The dominate channel is the $s_{\pm}$-wave pairing state, which possesses $s$-wave symmetry with sign reversal of the gap functions between the $\beta$ and $\alpha$ pockets, as well as between the $\beta$ and $\gamma$ pockets. As shown in Fig.~\ref{f1}(a), the gap function is positive on the $\beta$ pocket and negative on the $\alpha$ and $\gamma$ pockets. This is broadly consistent with previous studies~\cite{EreminRPAandpairing,YangfanPRLOdefi,rpazhangyangNC,rpaBraz,FRGwang,FRGwu,FRGwu2026}. It should be noted, however, that the gap function is anisotropic on the Fermi surfaces. In particular, on the $\beta$ pocket, the superconducting gap reaches its maximum magnitude around the $(0,\pi)$ and $(\pi, 0)$ regions, while exhibiting nodes along the diagonal direction. The subdominant channels are the $d_{xy}$-wave and $d_{x^2-y^2}$-wave pairing states, as shown in Fig.~\ref{f1}(b) and (c), respectively. The gap functions exhibit the standard $d_{xy}$ or $d_{x^2-y^2}$ symmetries on the Fermi surfaces. The three pairing channels are found to be nearly degenerate at small $U$. As $U$ increases, their pairing strengths all grow steadily, but the $s_{\pm}$ channel grows more rapidly than the $d_{xy}$ and $d_{x^2-y^2}$ channels. Consequently, it dominates the superconductivity at large $U$. 

\begin{figure}
	\centering\includegraphics[width=0.45\textwidth]{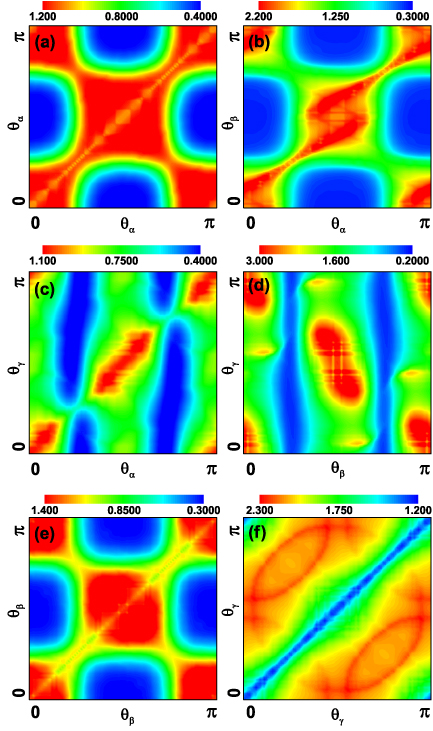}
	\caption{(Color online) 
		Fermi-angle dependence of the pairing interactions for $U=1.2$ and $J=U/6$ at a filling of $n=3.0$. (a)-(f) $\theta$-space distributions of $\Gamma_{\alpha\alpha}$, $\Gamma_{\alpha\beta}$, $\Gamma_{\alpha\gamma}$, $\Gamma_{\beta\gamma}$, $\Gamma_{\beta\beta}$ and $\Gamma_{\gamma\gamma}$, respectively.}
	\label{f2}
\end{figure}

To understand the origin of the $s_{\pm}$-wave pairing, we show in Fig.~\ref{f2} the $\theta$-dependence of the pairing interactions on the Fermi surfaces. Here, $\theta$ is the angle between the Fermi momentum and the horizontal axis, as illustrated in Fig.~\ref{f1}(a). Clear patterns emerge for the pairing interactions $\Gamma_{\mu\nu}(k,k')$ in $\theta$-space, where the peaked regions of $\Gamma_{\mu\nu}(k,k')$ denote the dominate pairing interactions. These regions are related to one another by translations of $(\pi/2,\pi/2)$, $(0,\pi)$ and their compositions in $\theta$-space. This can be understood by considering the symmetry properties and the spin-singlet nature of the pairing. The system possesses $C_4$ rotational symmetry, which shifts $\theta$ by $\pi/2$ and thus gives rise to a $(\pi/2,\pi/2)$ translation in $\theta$-space. On the other hand, owing to the spin-singlet nature, the pairing interactions satisfy $\Gamma_{\mu\nu}(k,k')=\Gamma_{\mu\nu}(k,-k')$. Since the momenta $k'$ and $-k'$ differ in angle by $\pi$, this gives rise to the $(0,\pi)$ displacement of the peaked regions. Thus, all peaked regions in $\theta$-space are connected by $(\pi/2,\pi/2)$, $(0,\pi)$ translations and their compositions, as a direct consequence of $C_4$ rotational symmetry and the spin-singlet pairing. Most importantly, the features of the pairing interactions in $\theta$-space can be exploited to identify the magnetic modes responsible for the emergence of superconductivity. 

As shown in Fig.~\ref{f2}, it is evident that $\Gamma_{\alpha\beta}$, $\Gamma_{\beta\gamma}$  and $\Gamma_{\gamma\gamma}$ are significantly larger than the other three pairing interactions among the six $\Gamma_{\mu\nu}$.  They therefore  dominate the pairing behavior of the system. For $\Gamma_{\alpha\beta}$, as depicted in Fig.~\ref{f1}(a), the peaks around $(\pi/2,\pi/2)$ in $\theta$-space stem from the $q_{2}^{*}$ magnetic excitations originating from the particle-hole scatterings between the $\alpha$ and $\beta$ pockets. As revealed by previous studies~\cite{zhangnjp,Botzel}, the spin excitations in the bilayer nickelate exhibit pronounced vertical momentum dependence. They are dominated by the magnetic odd modes with vertical momentum $\pi$, reflecting the antiferromagnetic correlation between spins in the upper and lower layers. To clarify the property of the pairing interaction, we note that the pairing vertex, which is dominated by spin fluctuations, can be expressed as $V^{p}_{\mu\nu} \sim [\hat{U}^{S}\hat{\mathcal{X}}^{S}\hat{U}^{S}]_{im,jl}\tilde{\phi}_{\mu\nu}(im)\tilde{\phi}_{\mu\nu}(jl)$, where $\tilde{\phi}_{\mu\nu}(im)=\phi^{*}_{\mu}(i)\phi_{\nu}(m)$ is the product of Bloch wavefunctions, and the summation over $i$, $j$, $l$ and $m$ is implied. To reveal the layer-resolved physics, we retain only the layer indices and suppress all others. The above expression then reduces to
\begin{align}
&[\hat{U}^{S}\hat{\mathcal{X}}^{S}\hat{U}^{S}]_{LL,L'L'}\tilde{\phi}_{\mu\nu}(LL)\tilde{\phi}_{\mu\nu}(L'L') \nonumber \\
=&\hat{U}^{S}\hat{\mathcal{X}}^{S}_{LL,L'L'}\hat{U}^{S}\tilde{\phi}_{\mu\nu}(LL)\tilde{\phi}_{\mu\nu}(L'L') \nonumber \\
=&\hat{U}^{S}\hat{\mathcal{X}}^{S}_{LL,LL}\hat{U}^{S}\tilde{\phi}_{\mu\nu}(LL)\tilde{\phi}_{\mu\nu}(LL)+ \nonumber \\
&\hat{U}^{S}\hat{\mathcal{X}}^{S}_{LL,\bar{L}\bar{L}}\hat{U}^{S}\tilde{\phi}_{\mu\nu}(LL)\tilde{\phi}_{\mu\nu}(\bar{L}\bar{L}) \nonumber \\
=&\hat{U}^{S}(\hat{\mathcal{X}}^{S}_{LL,LL}+P_{\mu}P_{\nu}\hat{\mathcal{X}}^{S}_{LL,\bar{L}\bar{L}})\hat{U}^{S}\tilde{\phi}_{\mu\nu}(LL)\tilde{\phi}_{\mu\nu}(LL)
\label{eq2}
\end{align}
In lines $1$ and $2$, we have used the fact that $\hat{U}^{S}$ is layer-diagonal and layer-independent, as only onsite interactions are considered in this work. In lines $3$ and $4$, the spin-fluctuation-mediated interaction is decomposed into intralayer and interlayer contributions. In line $5$, we have applied the relation $\tilde{\phi}_{\mu\nu}(\bar{L}\bar{L})=P_{\mu}P_{\nu}\tilde{\phi}_{\mu\nu}(LL)$, where $P_{\mu}$ and $P_{\nu}$ denote the parity of bands $\mu$ and $\nu$, respectively. As indicated by equation~\eqref{eq2}, the pair scatterings between bands of opposite parity arise from magnetic odd modes, while those between or within bands of the same parity stem from magnetic even modes. This finding aligns with the established picture that particle-hole excitations across opposite-parity bands produce magnetic odd modes, while same-parity excitations produce magnetic even modes~\cite{zhangnjp}. Given that $P_{\alpha}=1$ and $P_{\beta}=-1$, equation~\eqref{eq2} implies that $\Gamma_{\alpha\beta}$ derives entirely from the odd modes of the magnetic excitations. Consequently, the pronounced peak of $\Gamma_{\alpha\beta}$ around $(\pi/2,\pi/2)$ can be attributed to the $q_{2}^{*}$ magnetic odd modes. As emphasized earlier, the remaining peak structures are related to the $(\pi/2,\pi/2)$ region by $C_{4}$ rotational symmetry and the spin-singlet nature of the pairing. The wavevector $q_{2}^{*}$ and its symmetry-related counterparts connect all parallel segments of the $\alpha$ and $\beta$ pockets. Thus, as indicated in Fig.~\ref{f1}(a), these strong repulsions favor sign reversal of the gap functions between those segments. In this way, we establish a precise link between the magnetic modes and the pairing behavior -- a connection that has remained unclear in previous studies. $\Gamma_{\beta\gamma}$ is dominated by the peak around $(0,\pi)$ and its symmetry-related partners. In analogy to the $\Gamma_{\alpha\beta}$ case, these peaks can be attributed to the $q_{1}^{*}$ magnetic odd modes, which originate from particle-hole scatterings between the $\beta$ and $\gamma$ pockets, given that $P_{\beta}=-1$ and $P_{\gamma}=1$. Similarly, these peaks in $\Gamma_{\beta\gamma}$ drive sign reversal of the gap functions between the parallel segments of the $\beta$ and $\gamma$ pockets. Strong repulsion also exists  within the $\gamma$ pocket. However, as shown in Fig.~\ref{f2}, $\Gamma_{\alpha\beta}$ and $\Gamma_{\beta\gamma}$ dominate the pairing. Hence, the gap functions reverse their sign between the parallel segments of the $\alpha$ and $\beta$ pockets, as well as between those of the $\beta$ and $\gamma$ pockets. The $s_{\pm}$-wave pairing state is therefore the most energetically favorable. To take full advantage of the strong interpocket repulsion, the gap functions attain their largest magnitude around these segments of the Fermi pockets. Interestingly, the superconducting gaps are highly anisotropic on the $\alpha$ and $\beta$ Fermi pockets, with the largest magnitude occurring around the flat portions. However, gap nodes may appear around the diagonal region of these pockets. This stands in sharp contrast to the experimental data on bilayer nickelate thin films, where the Fermi pockets are found to be fully gapped~\cite{Chen2026film,NieYF2026NPfilm}. We note that, as shown in Fig.~\ref{f1}, both the  $d_{xy}$ and $d_{x^2-y^2}$ pairing states are also unconventional. They at least partially satisfy the established sign-reversal rules for the parallel segments of the Fermi pockets. The system thus hosts multiple competing pairing channels. At small $U$, these channels are nearly degenerate. However, the $s_{\pm}$ channel prevails over other competing ones as $U$ increases, owing to the fact that it takes full advantage of the interpocket repulsion in $\theta$-space.

\subsection{Doping behaviors of the pairing symmetry around $n=3.0$}

\begin{figure}
	\centering\includegraphics[width=0.45\textwidth]{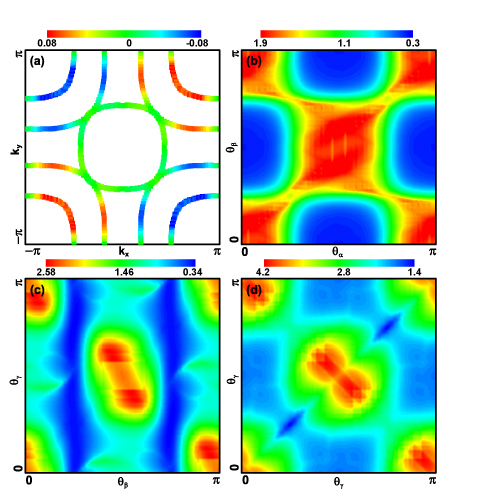}
	\caption{(Color online) 
		Superconducting gap and pairing interaction distributions for $U=1.2$ and $J=U/6$ at a filling of $n=2.7$. (a) Superconducting gap on the Fermi surfaces. (b)-(d) $\theta$-space distributions of  $\Gamma_{\alpha\beta}$, $\Gamma_{\beta\gamma}$ and $\Gamma_{\gamma\gamma}$, respectively.  }
	\label{f3}
\end{figure}

We further investigate the doping dependence of the pairing symmetry in pressurized La$_3$Ni$_2$O$_7$, focusing first on the hole-doped scenario. Owing to the much smaller Fermi velocity of the $\gamma$ pocket compared to those of the $\alpha$ and $\beta$ pockets, it expands rapidly in size with hole doping, whereas the $\alpha$ and $\beta$ pockets change only slightly. We find that the pairing interactions remain dominated by $\Gamma_{\gamma\gamma}$, $\Gamma_{\alpha\beta}$ and $\Gamma_{\beta\gamma}$. However, as the $\gamma$ pocket grows progressively larger, well-defined intrapocket nesting gradually emerges ~\cite{zhangnjp,rpaLlfDagtoMagPhas}. Consequently, new magnetic even modes develop from particle-hole excitations within the $\gamma$ pocket, strongly enhancing the intrapocket pairing interaction on this pocket, as shown in Fig.~\ref{f3}(d). This strong intrapocket repulsion tends to induce a sign reversal of gap function within the $\gamma$ pocket itself, while the sign-reversal rules driven from $\Gamma_{\alpha\beta}$ and $\Gamma_{\beta\gamma}$ continue to govern the interpocket physics. A typical pairing pattern in the heavily hole-doped regime at $n=2.7$ is shown in Fig.~\ref{f3}, where the dominate pairing symmetry is clearly $d_{xy}$-wave. We find this pairing symmetry to be quite general and insensitive to the interaction parameters, which can be attributed to the enhanced intrapocket interaction. Thus, a transition in pairing symmetry from $s_{\pm}$-wave to $d_{xy}$-wave occurs upon hole doping. This transition is driven by the strong enhancement of the magnetic even modes induced by hole doping. We anticipate this hole-doping-driven effect on the pairing state will be explored in future experimental investigations.

\begin{figure}
	\centering\includegraphics[width=0.45\textwidth]{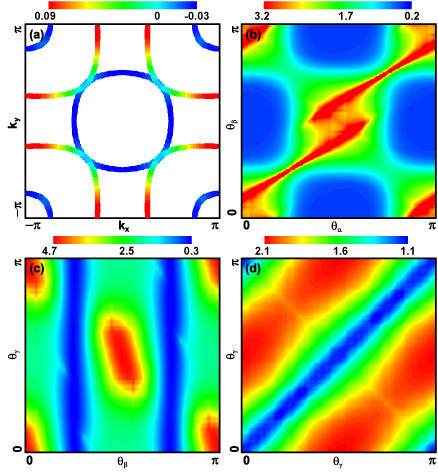}
	\caption{(Color online) 
		Superconducting gap and pairing interaction distributions for $U=1.2$ and $J=U/6$ at a filling of $n=3.3$. (a) Superconducting gap on the Fermi surfaces. (b)-(d) $\theta$-space distributions of  $\Gamma_{\alpha\beta}$, $\Gamma_{\beta\gamma}$ and $\Gamma_{\gamma\gamma}$, respectively.}
	\label{f4}
\end{figure}

Electron doping, on the other hand, expands the $\alpha$ pocket, while the hole $\beta$ and $\gamma$ pockets shrink gradually. Owing to the flat-band effect, the $\gamma$ pocket shrinks more rapidly under electron doping. Thus, a Lifshitz transition is expected in the heavily electron-doped regime. It has been emphasized that the emergence of the $\gamma$ pocket near the Fermi level is necessary for superconductivity~\cite{SunNat,EreminRPAandpairing,YangfanPRLOdefi,rpazhangyangNC,rpaBraz,FRGwang,FRGwu,FRGwu2026,FLEXdvs}. We therefore carefully studied the evolution of the pairing symmetry with electron doping. The interpocket interaction $\Gamma_{\beta\gamma}$ is found to remain dominant, despite the shrinkage of the $\gamma$ pocket as the system approaches the Lifshitz transition. Meanwhile, $\Gamma_{\alpha\beta}$ increases steadily with electron doping. As revealed by the previous study~\cite{zhangnjp}, the $\alpha$ and $\beta$ pockets begin to exhibit enhanced nesting behavior around the Lifshitz transition point, owing to their nearly equal sizes and similar Fermi surface topology. Consequently, as shown in Fig.~\ref{f4}, the strong repulsive interaction $\Gamma_{\alpha\beta}$ induces sign reversal of the gap functions between the  $\alpha$ and $\beta$ pockets. In parallel, the repulsion $\Gamma_{\beta\gamma}$ drives opposite signs of the gap functions on the $\beta$ and $\gamma$ pockets. Therefore, the $s_{\pm}$-wave pairing persists into the electron-doped regime prior to the Lifshitz transition, even as the $\gamma$ pocket gradually vanishes with electron doping. Thus, it seems that the $\gamma$ pocket is not essential for the emergence of superconductivity in carrier-doped pressurized La$_3$Ni$_2$O$_7$, in contrary with the proposal that the $\gamma$ Fermi pocket is a prerequisite for superconductivity in bilayer nickelates~\cite{SunNat,EreminRPAandpairing,YangfanPRLOdefi,rpazhangyangNC,rpaBraz,FRGwang,FRGwu,FRGwu2026,FLEXdvs}. In fact, the absence of the $\gamma$ pocket is accompanied by an enhancement of the magnetic odd $q_{2}^{*}$ modes, which stems from the improved  nesting between the $\alpha$ and $\beta$ Fermi pockets under electron doping.

\begin{figure}
	\centering\includegraphics[width=0.45\textwidth]{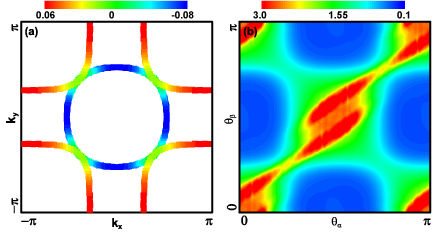}
	\caption{(Color online) 
		Superconducting gap and pairing interaction distributions for $U=1.1$ and $J=U/6$ at a filling of $n=3.5$. (a) Superconducting gap on the Fermi surfaces. (b) $\theta$-space distributions of  $\Gamma_{\alpha\beta}$.}
	\label{f5}
\end{figure}

\subsection{The pairing symmetry below the Lifshitz transition}
With further electron doping, the $\gamma$ pocket sinks below the Fermi level, i.e., a Lifshitz transition occurs around $n=3.4$. This provides a playground for exploring the role of the $\gamma$ pocket in the superconducting pairing of pressurized bilayer nickelates. The typical results are illustrated in Fig.~\ref{f5}. The pairing interactions are found to be dominated by $\Gamma_{\alpha\beta}$. It can be seen that the leading pairing state remains the $s_{\pm}$-wave symmetry, with sign reversal of the gap functions between the $\alpha$ and $\beta$ pockets, even though the $\gamma$ pocket lies entirely below the Fermi level. As mentioned above, this $s_{\pm}$-pairing stems from the magnetic odd $q_{2}^{*}$ modes.

The persistence of the $s_{\pm}$-wave pairing state below the Lifshitz transition can be understood as follows. In the absence of the $\gamma$ Fermi pocket, it is fully occupied with an electron count of $n_{\gamma}=2$. According to Luttinger's theorem~\cite{Luttingertheorem}, the sizes of the $\alpha$ and $\beta$ pockets must then be equal, satisfying $n_{\alpha}+n_{\beta}=2$, provided the total filling of the system is $n=4$. Given that both Fermi pockets exhibit a square-like shape, perfect nesting between them can be expected, similar to the scenario of the iron-based superconductors~\cite{Mazin}. Consequently, strong spin fluctuations should emerge as the filling approaches $n=4$. As a result, the interpocket repulsion can be significantly enhanced below the Lifshitz transition. In Fig.~\ref{f5}(b), we show $\Gamma_{\alpha\beta}$ which is substantially larger than the intrapocket ones (not shown here). This strong repulsion induces sign reversal of the gap functions between the $\alpha$ and $\beta$ pockets, giving rise to the unusual $s_{\pm}$-wave pairing state. Crucially, this pairing state develops with the $\gamma$ band lying entirely below the Fermi level. This further confirms that the $\gamma$ Fermi pocket is not essential for superconductivity in pressurized bilayer nickelates, in contrast to the proposal put forward by a recent experimental study~\cite{Nie2026film}. This theoretical prediction calls for experimental verification in future studies. We note that the situation here differs markedly from that of nickelate thin films, where the filling is around $n=3.0$. Nevertheless, we emphasize that superconductivity can emerge even in the absence of the $\gamma$ Fermi pocket, provided good nesting exists between the $\alpha$ and $\beta$ pockets. On the other hand, it should be noted that competition may arise between $s_{\pm}$-wave superconductivity and the SDW order below the Lifshitz transition. We propose that comprehensive investigations of both bilayer nickelate thin films and carrier-doped La$_3$Ni$_2$O$_7$ under pressure may help unravel the mechanism of superconductivity.

\section{Summary and conclusions}
In summary, we have investigated the doping evolution of the superconducting pairing in pressurized La$_3$Ni$_2$O$_7$. The leading pairing symmetry is found to be the $s_{\pm}$-wave type for the undoped compound. A detailed analysis of the pairing interactions reveals that they are dominated by strong repulsions between the $\alpha$ and $\beta$ pockets, and between the $\beta$ and $\gamma$ pockets. These repulsions arise from the magnetic odd $q_{1}^{*}$ and $q_{2}^{*}$ modes, which originate from the particle-hole excitations between bands of opposite parities~\cite{zhangnjp}. As a result, the most favorable pairing state is the $s_{\pm}$-wave state, which exhibits sign reversal of gap functions between the $\beta$ and $\alpha$ pockets, as well as between $\beta$ and $\gamma$ pockets. A key finding of the present work is thus the identification of magnetic odd modes as the microscopic mechanism responsible for the widely discussed $s_{\pm}$-wave pairing in pressurized La$_3$Ni$_2$O$_7$ --- a connection that has remained unclear in previous theoretical studies~\cite{EreminRPAandpairing,YangfanPRLOdefi,rpazhangyangNC,rpaBraz,FRGwang,FRGwu,FRGwu2026}. 

Upon hole doping, the expansion of the $\gamma$ pocket enhances intrapocket repulsion, i.e., pair scatterings within the $\gamma$ pocket become significant. This repulsive interaction arises from the magnetic even modes closely related to the intrapocket particle-hole excitations. Consequently, it drives a transition in pairing symmetry from $s_{\pm}$-wave to $d_{xy}$-wave in the heavily hole-doped regime. This transition originates from the shift of magnetic excitations from odd-mode-dominant to even-mode-dominant with increasing hole doping. For the electron-doped scenario, we find that $\Gamma_{\alpha\beta}$ increases steadily, owing to the favorable nesting behavior of the $\alpha$ and $\beta$ pockets induced by electron doping. Meanwhile, $\Gamma_{\beta\gamma}$ remains substantial. Hence, the $s_{\pm}$-wave pairing state persists into the heavily electron-doped regime prior to the Lifshitz transition. It is further found that, below the Lifshitz transition, the pairing interaction $\Gamma_{\alpha\beta}$ is strongly enhanced owing to the improved nesting between the $\alpha$ and $\beta$ pockets, which in turn results from the disappearance of the $\gamma$ Fermi pocket. Thus, $s_{\pm}$-wave pairing persists below the Lifshitz transition. The existence of the $\gamma$ Fermi pocket, therefore, is not a prerequisite for superconductivity in pressurized bilayer nickelates.

Although the doping evolution of the pairing symmetry is established for pressurized bilayer nickelates in the present work, we note that previous studies have indicated that the pairing symmetry is sensitive to the low-energy band structures~\cite{bands31,CrystalNC}. Meanwhile, nonlocal interactions have been shown to drive the transition of the pairing symmetry from $s$-wave to $d$-wave~\cite{InterlayLi,rpaBraz,FRGwu2026}. Therefore, to establish a realistic theoretical model for bilayer nickelates under high pressure, both high-resolution electronic band structure data and experimental verification of the proposed pairing symmetries are urgently needed.

\section{Acknowledgments}
This work was supported by the National Natural Science Foundation of China (Grants Nos.~11804290, 11647072).


\begin{thebibliography}{56}%
	\makeatletter
	\providecommand \@ifxundefined [1]{%
		\@ifx{#1\undefined}
	}%
	\providecommand \@ifnum [1]{%
		\ifnum #1\expandafter \@firstoftwo
		\else \expandafter \@secondoftwo
		\fi
	}%
	\providecommand \@ifx [1]{%
		\ifx #1\expandafter \@firstoftwo
		\else \expandafter \@secondoftwo
		\fi
	}%
	\providecommand \natexlab [1]{#1}%
	\providecommand \enquote  [1]{``#1''}%
	\providecommand \bibnamefont  [1]{#1}%
	\providecommand \bibfnamefont [1]{#1}%
	\providecommand \citenamefont [1]{#1}%
	\providecommand \href@noop [0]{\@secondoftwo}%
	\providecommand \href [0]{\begingroup \@sanitize@url \@href}%
	\providecommand \@href[1]{\@@startlink{#1}\@@href}%
	\providecommand \@@href[1]{\endgroup#1\@@endlink}%
	\providecommand \@sanitize@url [0]{\catcode `\\12\catcode `\$12\catcode
		`\&12\catcode `\#12\catcode `\^12\catcode `\_12\catcode `\%12\relax}%
	\providecommand \@@startlink[1]{}%
	\providecommand \@@endlink[0]{}%
	\providecommand \url  [0]{\begingroup\@sanitize@url \@url }%
	\providecommand \@url [1]{\endgroup\@href {#1}{\urlprefix }}%
	\providecommand \urlprefix  [0]{URL }%
	\providecommand \Eprint [0]{\href }%
	\providecommand \doibase [0]{https://doi.org/}%
	\providecommand \selectlanguage [0]{\@gobble}%
	\providecommand \bibinfo  [0]{\@secondoftwo}%
	\providecommand \bibfield  [0]{\@secondoftwo}%
	\providecommand \translation [1]{[#1]}%
	\providecommand \BibitemOpen [0]{}%
	\providecommand \bibitemStop [0]{}%
	\providecommand \bibitemNoStop [0]{.\EOS\space}%
	\providecommand \EOS [0]{\spacefactor3000\relax}%
	\providecommand \BibitemShut  [1]{\csname bibitem#1\endcsname}%
	\let\auto@bib@innerbib\@empty
	\bibitem [{\citenamefont {Sun}\ \emph {et~al.}(2023)\citenamefont {Sun},
		\citenamefont {Huo}, \citenamefont {Hu}, \citenamefont {Li}, \citenamefont
		{Liu}, \citenamefont {Han}, \citenamefont {Tang}, \citenamefont {Mao},
		\citenamefont {Yang}, \citenamefont {Wang}, \citenamefont {Cheng},
		\citenamefont {Yao}, \citenamefont {Zhang},\ and\ \citenamefont
		{Wang}}]{SunNat}%
	\BibitemOpen
	\bibfield  {author} {\bibinfo {author} {\bibfnamefont {H.}~\bibnamefont
			{Sun}}, \bibinfo {author} {\bibfnamefont {M.}~\bibnamefont {Huo}}, \bibinfo
		{author} {\bibfnamefont {X.}~\bibnamefont {Hu}}, \bibinfo {author}
		{\bibfnamefont {J.}~\bibnamefont {Li}}, \bibinfo {author} {\bibfnamefont
			{Z.}~\bibnamefont {Liu}}, \bibinfo {author} {\bibfnamefont {Y.}~\bibnamefont
			{Han}}, \bibinfo {author} {\bibfnamefont {L.}~\bibnamefont {Tang}}, \bibinfo
		{author} {\bibfnamefont {Z.}~\bibnamefont {Mao}}, \bibinfo {author}
		{\bibfnamefont {P.}~\bibnamefont {Yang}}, \bibinfo {author} {\bibfnamefont
			{B.}~\bibnamefont {Wang}}, \bibinfo {author} {\bibfnamefont {J.}~\bibnamefont
			{Cheng}}, \bibinfo {author} {\bibfnamefont {D.-X.}\ \bibnamefont {Yao}},
		\bibinfo {author} {\bibfnamefont {G.-M.}\ \bibnamefont {Zhang}},\ and\
		\bibinfo {author} {\bibfnamefont {M.}~\bibnamefont {Wang}},\ }\bibfield
	{title} {\bibinfo {title} {Signatures of superconductivity near 80閳ュ
			in a nickelate under high pressure},\ }\href
	{https://doi.org/10.1038/s41586-023-06408-7} {\bibfield  {journal} {\bibinfo
			{journal} {Nature}\ }\textbf {\bibinfo {volume} {621}},\ \bibinfo {pages}
		{493} (\bibinfo {year} {2023})}\BibitemShut {NoStop}%
	\bibitem [{\citenamefont {Zhou}\ \emph {et~al.}(2025)\citenamefont {Zhou},
		\citenamefont {Guo}, \citenamefont {Cai}, \citenamefont {Sun}, \citenamefont
		{Li}, \citenamefont {Zhao}, \citenamefont {Wang}, \citenamefont {Han},
		\citenamefont {Chen}, \citenamefont {Chen}, \citenamefont {Wu}, \citenamefont
		{Ding}, \citenamefont {Xiang}, \citenamefont {Mao},\ and\ \citenamefont
		{Sun}}]{zerofilamentary}%
	\BibitemOpen
	\bibfield  {author} {\bibinfo {author} {\bibfnamefont {Y.}~\bibnamefont
			{Zhou}}, \bibinfo {author} {\bibfnamefont {J.}~\bibnamefont {Guo}}, \bibinfo
		{author} {\bibfnamefont {S.}~\bibnamefont {Cai}}, \bibinfo {author}
		{\bibfnamefont {H.}~\bibnamefont {Sun}}, \bibinfo {author} {\bibfnamefont
			{C.}~\bibnamefont {Li}}, \bibinfo {author} {\bibfnamefont {J.}~\bibnamefont
			{Zhao}}, \bibinfo {author} {\bibfnamefont {P.}~\bibnamefont {Wang}}, \bibinfo
		{author} {\bibfnamefont {J.}~\bibnamefont {Han}}, \bibinfo {author}
		{\bibfnamefont {X.}~\bibnamefont {Chen}}, \bibinfo {author} {\bibfnamefont
			{Y.}~\bibnamefont {Chen}}, \bibinfo {author} {\bibfnamefont {Q.}~\bibnamefont
			{Wu}}, \bibinfo {author} {\bibfnamefont {Y.}~\bibnamefont {Ding}}, \bibinfo
		{author} {\bibfnamefont {T.}~\bibnamefont {Xiang}}, \bibinfo {author}
		{\bibfnamefont {H.-k.}\ \bibnamefont {Mao}},\ and\ \bibinfo {author}
		{\bibfnamefont {L.}~\bibnamefont {Sun}},\ }\bibfield  {title} {\bibinfo
		{title} {Investigations of key issues on the reproducibility of high-tc
			superconductivity emerging from compressed la3ni2o7},\ }\href
	{https://doi.org/10.1063/5.0247684} {\bibfield  {journal} {\bibinfo
			{journal} {Matter Radiat. Extremes}\ }\textbf {\bibinfo {volume} {10}},\
		\bibinfo {pages} {027801} (\bibinfo {year} {2025})}\BibitemShut {NoStop}%
	\bibitem [{\citenamefont {Zhang}\ \emph
		{et~al.}(2024{\natexlab{a}})\citenamefont {Zhang}, \citenamefont {Su},
		\citenamefont {Huang}, \citenamefont {Shan}, \citenamefont {Sun},
		\citenamefont {Huo}, \citenamefont {Ye}, \citenamefont {Zhang}, \citenamefont
		{Yang}, \citenamefont {Xu}, \citenamefont {Su}, \citenamefont {Li},
		\citenamefont {Smidman}, \citenamefont {Wang}, \citenamefont {Jiao},\ and\
		\citenamefont {Yuan}}]{zeroR1}%
	\BibitemOpen
	\bibfield  {author} {\bibinfo {author} {\bibfnamefont {Y.}~\bibnamefont
			{Zhang}}, \bibinfo {author} {\bibfnamefont {D.}~\bibnamefont {Su}}, \bibinfo
		{author} {\bibfnamefont {Y.}~\bibnamefont {Huang}}, \bibinfo {author}
		{\bibfnamefont {Z.}~\bibnamefont {Shan}}, \bibinfo {author} {\bibfnamefont
			{H.}~\bibnamefont {Sun}}, \bibinfo {author} {\bibfnamefont {M.}~\bibnamefont
			{Huo}}, \bibinfo {author} {\bibfnamefont {K.}~\bibnamefont {Ye}}, \bibinfo
		{author} {\bibfnamefont {J.}~\bibnamefont {Zhang}}, \bibinfo {author}
		{\bibfnamefont {Z.}~\bibnamefont {Yang}}, \bibinfo {author} {\bibfnamefont
			{Y.}~\bibnamefont {Xu}}, \bibinfo {author} {\bibfnamefont {Y.}~\bibnamefont
			{Su}}, \bibinfo {author} {\bibfnamefont {R.}~\bibnamefont {Li}}, \bibinfo
		{author} {\bibfnamefont {M.}~\bibnamefont {Smidman}}, \bibinfo {author}
		{\bibfnamefont {M.}~\bibnamefont {Wang}}, \bibinfo {author} {\bibfnamefont
			{L.}~\bibnamefont {Jiao}},\ and\ \bibinfo {author} {\bibfnamefont
			{H.}~\bibnamefont {Yuan}},\ }\bibfield  {title} {\bibinfo {title}
		{High-temperature superconductivity with zero resistance and strange-metal
			behaviour in ${\mathrm{la}}_{3}{\text{ni}}_{2}{\mathrm{o}}_{7-\delta}$},\
	}\href {https://doi.org/10.1038/s41567-024-02515-y} {\bibfield  {journal}
		{\bibinfo  {journal} {Nature Physics}\ }\textbf {\bibinfo {volume} {20}},\
		\bibinfo {pages} {1269} (\bibinfo {year} {2024}{\natexlab{a}})}\BibitemShut
	{NoStop}%
	\bibitem [{\citenamefont {Hou}\ \emph {et~al.}(2023)\citenamefont {Hou},
		\citenamefont {Yang}, \citenamefont {Liu}, \citenamefont {Li}, \citenamefont
		{Shan}, \citenamefont {Ma}, \citenamefont {Wang}, \citenamefont {Wang},
		\citenamefont {Guo}, \citenamefont {Sun}, \citenamefont {Uwatoko},
		\citenamefont {Wang}, \citenamefont {Zhang}, \citenamefont {Wang},\ and\
		\citenamefont {Cheng}}]{zeroR2}%
	\BibitemOpen
	\bibfield  {author} {\bibinfo {author} {\bibfnamefont {J.}~\bibnamefont
			{Hou}}, \bibinfo {author} {\bibfnamefont {P.-T.}\ \bibnamefont {Yang}},
		\bibinfo {author} {\bibfnamefont {Z.-Y.}\ \bibnamefont {Liu}}, \bibinfo
		{author} {\bibfnamefont {J.-Y.}\ \bibnamefont {Li}}, \bibinfo {author}
		{\bibfnamefont {P.-F.}\ \bibnamefont {Shan}}, \bibinfo {author}
		{\bibfnamefont {L.}~\bibnamefont {Ma}}, \bibinfo {author} {\bibfnamefont
			{G.}~\bibnamefont {Wang}}, \bibinfo {author} {\bibfnamefont {N.-N.}\
			\bibnamefont {Wang}}, \bibinfo {author} {\bibfnamefont {H.-Z.}\ \bibnamefont
			{Guo}}, \bibinfo {author} {\bibfnamefont {J.-P.}\ \bibnamefont {Sun}},
		\bibinfo {author} {\bibfnamefont {Y.}~\bibnamefont {Uwatoko}}, \bibinfo
		{author} {\bibfnamefont {M.}~\bibnamefont {Wang}}, \bibinfo {author}
		{\bibfnamefont {G.-M.}\ \bibnamefont {Zhang}}, \bibinfo {author}
		{\bibfnamefont {B.-S.}\ \bibnamefont {Wang}},\ and\ \bibinfo {author}
		{\bibfnamefont {J.-G.}\ \bibnamefont {Cheng}},\ }\bibfield  {title} {\bibinfo
		{title} {Emergence of high-temperature superconducting phase in pressurized
			la3ni2o7 crystals},\ }\href {https://doi.org/10.1088/0256-307X/40/11/117302}
	{\bibfield  {journal} {\bibinfo  {journal} {Chinese Physics Letters}\
		}\textbf {\bibinfo {volume} {40}},\ \bibinfo {pages} {117302} (\bibinfo
		{year} {2023})}\BibitemShut {NoStop}%
	\bibitem [{\citenamefont {Li}\ \emph {et~al.}(2025{\natexlab{a}})\citenamefont
		{Li}, \citenamefont {Peng}, \citenamefont {Ma}, \citenamefont {Zhang},
		\citenamefont {Xing}, \citenamefont {Huang}, \citenamefont {Huang},
		\citenamefont {Huo}, \citenamefont {Hu}, \citenamefont {Dong}, \citenamefont
		{Chen}, \citenamefont {Xie}, \citenamefont {Dong}, \citenamefont {Sun},
		\citenamefont {Zeng}, \citenamefont {Mao},\ and\ \citenamefont
		{Wang}}]{zeroR4}%
	\BibitemOpen
	\bibfield  {author} {\bibinfo {author} {\bibfnamefont {J.}~\bibnamefont
			{Li}}, \bibinfo {author} {\bibfnamefont {D.}~\bibnamefont {Peng}}, \bibinfo
		{author} {\bibfnamefont {P.}~\bibnamefont {Ma}}, \bibinfo {author}
		{\bibfnamefont {H.}~\bibnamefont {Zhang}}, \bibinfo {author} {\bibfnamefont
			{Z.}~\bibnamefont {Xing}}, \bibinfo {author} {\bibfnamefont {X.}~\bibnamefont
			{Huang}}, \bibinfo {author} {\bibfnamefont {C.}~\bibnamefont {Huang}},
		\bibinfo {author} {\bibfnamefont {M.}~\bibnamefont {Huo}}, \bibinfo {author}
		{\bibfnamefont {D.}~\bibnamefont {Hu}}, \bibinfo {author} {\bibfnamefont
			{Z.}~\bibnamefont {Dong}}, \bibinfo {author} {\bibfnamefont {X.}~\bibnamefont
			{Chen}}, \bibinfo {author} {\bibfnamefont {T.}~\bibnamefont {Xie}}, \bibinfo
		{author} {\bibfnamefont {H.}~\bibnamefont {Dong}}, \bibinfo {author}
		{\bibfnamefont {H.}~\bibnamefont {Sun}}, \bibinfo {author} {\bibfnamefont
			{Q.}~\bibnamefont {Zeng}}, \bibinfo {author} {\bibfnamefont {H.-k.}\
			\bibnamefont {Mao}},\ and\ \bibinfo {author} {\bibfnamefont {M.}~\bibnamefont
			{Wang}},\ }\bibfield  {title} {\bibinfo {title} {Identification of
			superconductivity in bilayer nickelate la3ni2o7 under high pressure up to 100
			gpa},\ }\href {https://doi.org/10.1093/nsr/nwaf220} {\bibfield  {journal}
		{\bibinfo  {journal} {Natl Sci Rev}\ }\textbf {\bibinfo {volume} {12}},\
		\bibinfo {pages} {nwaf220} (\bibinfo {year}
		{2025}{\natexlab{a}})}\BibitemShut {NoStop}%
	\bibitem [{\citenamefont {Wang}\ \emph
		{et~al.}(2024{\natexlab{a}})\citenamefont {Wang}, \citenamefont {Wang},
		\citenamefont {Shen}, \citenamefont {Hou}, \citenamefont {Luo}, \citenamefont
		{Ma}, \citenamefont {Yang}, \citenamefont {Shi}, \citenamefont {Dou},
		\citenamefont {Feng}, \citenamefont {Yang}, \citenamefont {Shi},
		\citenamefont {Ren}, \citenamefont {Ma}, \citenamefont {Yang}, \citenamefont
		{Liu}, \citenamefont {Liu}, \citenamefont {Zhang}, \citenamefont {Dong},
		\citenamefont {Wang}, \citenamefont {Jiang}, \citenamefont {Hu},
		\citenamefont {Nagasaki}, \citenamefont {Kitagawa}, \citenamefont {Calder},
		\citenamefont {Yan}, \citenamefont {Sun}, \citenamefont {Wang}, \citenamefont
		{Zhou}, \citenamefont {Uwatoko},\ and\ \citenamefont {Cheng}}]{zeroRandMesn}%
	\BibitemOpen
	\bibfield  {author} {\bibinfo {author} {\bibfnamefont {N.}~\bibnamefont
			{Wang}}, \bibinfo {author} {\bibfnamefont {G.}~\bibnamefont {Wang}}, \bibinfo
		{author} {\bibfnamefont {X.}~\bibnamefont {Shen}}, \bibinfo {author}
		{\bibfnamefont {J.}~\bibnamefont {Hou}}, \bibinfo {author} {\bibfnamefont
			{J.}~\bibnamefont {Luo}}, \bibinfo {author} {\bibfnamefont {X.}~\bibnamefont
			{Ma}}, \bibinfo {author} {\bibfnamefont {H.}~\bibnamefont {Yang}}, \bibinfo
		{author} {\bibfnamefont {L.}~\bibnamefont {Shi}}, \bibinfo {author}
		{\bibfnamefont {J.}~\bibnamefont {Dou}}, \bibinfo {author} {\bibfnamefont
			{J.}~\bibnamefont {Feng}}, \bibinfo {author} {\bibfnamefont {J.}~\bibnamefont
			{Yang}}, \bibinfo {author} {\bibfnamefont {Y.}~\bibnamefont {Shi}}, \bibinfo
		{author} {\bibfnamefont {Z.}~\bibnamefont {Ren}}, \bibinfo {author}
		{\bibfnamefont {H.}~\bibnamefont {Ma}}, \bibinfo {author} {\bibfnamefont
			{P.}~\bibnamefont {Yang}}, \bibinfo {author} {\bibfnamefont {Z.}~\bibnamefont
			{Liu}}, \bibinfo {author} {\bibfnamefont {Y.}~\bibnamefont {Liu}}, \bibinfo
		{author} {\bibfnamefont {H.}~\bibnamefont {Zhang}}, \bibinfo {author}
		{\bibfnamefont {X.}~\bibnamefont {Dong}}, \bibinfo {author} {\bibfnamefont
			{Y.}~\bibnamefont {Wang}}, \bibinfo {author} {\bibfnamefont {K.}~\bibnamefont
			{Jiang}}, \bibinfo {author} {\bibfnamefont {J.}~\bibnamefont {Hu}}, \bibinfo
		{author} {\bibfnamefont {S.}~\bibnamefont {Nagasaki}}, \bibinfo {author}
		{\bibfnamefont {K.}~\bibnamefont {Kitagawa}}, \bibinfo {author}
		{\bibfnamefont {S.}~\bibnamefont {Calder}}, \bibinfo {author} {\bibfnamefont
			{J.}~\bibnamefont {Yan}}, \bibinfo {author} {\bibfnamefont {J.}~\bibnamefont
			{Sun}}, \bibinfo {author} {\bibfnamefont {B.}~\bibnamefont {Wang}}, \bibinfo
		{author} {\bibfnamefont {R.}~\bibnamefont {Zhou}}, \bibinfo {author}
		{\bibfnamefont {Y.}~\bibnamefont {Uwatoko}},\ and\ \bibinfo {author}
		{\bibfnamefont {J.}~\bibnamefont {Cheng}},\ }\bibfield  {title} {\bibinfo
		{title} {Bulk high-temperature superconductivity in pressurized tetragonal
			la2prni2o7},\ }\href {https://doi.org/10.1038/s41586-024-07996-8} {\bibfield
		{journal} {\bibinfo  {journal} {Nature}\ }\textbf {\bibinfo {volume} {634}},\
		\bibinfo {pages} {579} (\bibinfo {year} {2024}{\natexlab{a}})}\BibitemShut
	{NoStop}%
	\bibitem [{\citenamefont {Wang}\ \emph
		{et~al.}(2024{\natexlab{b}})\citenamefont {Wang}, \citenamefont {Wang},
		\citenamefont {Shen}, \citenamefont {Hou}, \citenamefont {Ma}, \citenamefont
		{Shi}, \citenamefont {Ren}, \citenamefont {Gu}, \citenamefont {Ma},
		\citenamefont {Yang}, \citenamefont {Liu}, \citenamefont {Guo}, \citenamefont
		{Sun}, \citenamefont {Zhang}, \citenamefont {Calder}, \citenamefont {Yan},
		\citenamefont {Wang}, \citenamefont {Uwatoko},\ and\ \citenamefont
		{Cheng}}]{zeroRpolycrystal}%
	\BibitemOpen
	\bibfield  {author} {\bibinfo {author} {\bibfnamefont {G.}~\bibnamefont
			{Wang}}, \bibinfo {author} {\bibfnamefont {N.~N.}\ \bibnamefont {Wang}},
		\bibinfo {author} {\bibfnamefont {X.~L.}\ \bibnamefont {Shen}}, \bibinfo
		{author} {\bibfnamefont {J.}~\bibnamefont {Hou}}, \bibinfo {author}
		{\bibfnamefont {L.}~\bibnamefont {Ma}}, \bibinfo {author} {\bibfnamefont
			{L.~F.}\ \bibnamefont {Shi}}, \bibinfo {author} {\bibfnamefont {Z.~A.}\
			\bibnamefont {Ren}}, \bibinfo {author} {\bibfnamefont {Y.~D.}\ \bibnamefont
			{Gu}}, \bibinfo {author} {\bibfnamefont {H.~M.}\ \bibnamefont {Ma}}, \bibinfo
		{author} {\bibfnamefont {P.~T.}\ \bibnamefont {Yang}}, \bibinfo {author}
		{\bibfnamefont {Z.~Y.}\ \bibnamefont {Liu}}, \bibinfo {author} {\bibfnamefont
			{H.~Z.}\ \bibnamefont {Guo}}, \bibinfo {author} {\bibfnamefont {J.~P.}\
			\bibnamefont {Sun}}, \bibinfo {author} {\bibfnamefont {G.~M.}\ \bibnamefont
			{Zhang}}, \bibinfo {author} {\bibfnamefont {S.}~\bibnamefont {Calder}},
		\bibinfo {author} {\bibfnamefont {J.-Q.}\ \bibnamefont {Yan}}, \bibinfo
		{author} {\bibfnamefont {B.~S.}\ \bibnamefont {Wang}}, \bibinfo {author}
		{\bibfnamefont {Y.}~\bibnamefont {Uwatoko}},\ and\ \bibinfo {author}
		{\bibfnamefont {J.-G.}\ \bibnamefont {Cheng}},\ }\bibfield  {title} {\bibinfo
		{title} {{Pressure-Induced Superconductivity In Polycrystalline
				${\mathrm{La}}_{3}{\mathrm{Ni}}_{2}{\mathrm{O}}_{7\ensuremath{-}\ensuremath{\delta}}$}},\
	}\href {https://doi.org/10.1103/PhysRevX.14.011040} {\bibfield  {journal}
		{\bibinfo  {journal} {Phys. Rev. X}\ }\textbf {\bibinfo {volume} {14}},\
		\bibinfo {pages} {011040} (\bibinfo {year} {2024}{\natexlab{b}})}\BibitemShut
	{NoStop}%
	\bibitem [{\citenamefont {Li}\ \emph {et~al.}(2026)\citenamefont {Li},
		\citenamefont {Xing}, \citenamefont {Peng}, \citenamefont {Dou},
		\citenamefont {Guo}, \citenamefont {Ma}, \citenamefont {Zhang}, \citenamefont
		{Wang}, \citenamefont {Luo}, \citenamefont {Yang}, \citenamefont {Zhang},
		\citenamefont {Chang}, \citenamefont {Chen}, \citenamefont {Cai},
		\citenamefont {Cheng}, \citenamefont {Wang}, \citenamefont {Liu},
		\citenamefont {Luo}, \citenamefont {Hirao}, \citenamefont {Matsuoka},
		\citenamefont {Kadobayashi}, \citenamefont {Zeng}, \citenamefont {Zheng},
		\citenamefont {Zhou}, \citenamefont {Zeng}, \citenamefont {Tao},\ and\
		\citenamefont {Zhang}}]{bulk90K}%
	\BibitemOpen
	\bibfield  {author} {\bibinfo {author} {\bibfnamefont {F.}~\bibnamefont
			{Li}}, \bibinfo {author} {\bibfnamefont {Z.}~\bibnamefont {Xing}}, \bibinfo
		{author} {\bibfnamefont {D.}~\bibnamefont {Peng}}, \bibinfo {author}
		{\bibfnamefont {J.}~\bibnamefont {Dou}}, \bibinfo {author} {\bibfnamefont
			{N.}~\bibnamefont {Guo}}, \bibinfo {author} {\bibfnamefont {L.}~\bibnamefont
			{Ma}}, \bibinfo {author} {\bibfnamefont {Y.}~\bibnamefont {Zhang}}, \bibinfo
		{author} {\bibfnamefont {L.}~\bibnamefont {Wang}}, \bibinfo {author}
		{\bibfnamefont {J.}~\bibnamefont {Luo}}, \bibinfo {author} {\bibfnamefont
			{J.}~\bibnamefont {Yang}}, \bibinfo {author} {\bibfnamefont {J.}~\bibnamefont
			{Zhang}}, \bibinfo {author} {\bibfnamefont {T.}~\bibnamefont {Chang}},
		\bibinfo {author} {\bibfnamefont {Y.-S.}\ \bibnamefont {Chen}}, \bibinfo
		{author} {\bibfnamefont {W.}~\bibnamefont {Cai}}, \bibinfo {author}
		{\bibfnamefont {J.}~\bibnamefont {Cheng}}, \bibinfo {author} {\bibfnamefont
			{Y.}~\bibnamefont {Wang}}, \bibinfo {author} {\bibfnamefont {Y.}~\bibnamefont
			{Liu}}, \bibinfo {author} {\bibfnamefont {T.}~\bibnamefont {Luo}}, \bibinfo
		{author} {\bibfnamefont {N.}~\bibnamefont {Hirao}}, \bibinfo {author}
		{\bibfnamefont {T.}~\bibnamefont {Matsuoka}}, \bibinfo {author}
		{\bibfnamefont {H.}~\bibnamefont {Kadobayashi}}, \bibinfo {author}
		{\bibfnamefont {Z.}~\bibnamefont {Zeng}}, \bibinfo {author} {\bibfnamefont
			{Q.}~\bibnamefont {Zheng}}, \bibinfo {author} {\bibfnamefont
			{R.}~\bibnamefont {Zhou}}, \bibinfo {author} {\bibfnamefont {Q.}~\bibnamefont
			{Zeng}}, \bibinfo {author} {\bibfnamefont {X.}~\bibnamefont {Tao}},\ and\
		\bibinfo {author} {\bibfnamefont {J.}~\bibnamefont {Zhang}},\ }\bibfield
	{title} {\bibinfo {title} {Bulk superconductivity up to 96 k in pressurized
			nickelate single crystals},\ }\href
	{https://doi.org/10.1038/s41586-025-09954-4} {\bibfield  {journal} {\bibinfo
			{journal} {Nature}\ }\textbf {\bibinfo {volume} {649}},\ \bibinfo {pages}
		{871} (\bibinfo {year} {2026})}\BibitemShut {NoStop}%
	\bibitem [{\citenamefont {Damascelli}\ \emph {et~al.}(2003)\citenamefont
		{Damascelli}, \citenamefont {Hussain},\ and\ \citenamefont
		{Shen}}]{CupratesShenZhiXunRMP}%
	\BibitemOpen
	\bibfield  {author} {\bibinfo {author} {\bibfnamefont {A.}~\bibnamefont
			{Damascelli}}, \bibinfo {author} {\bibfnamefont {Z.}~\bibnamefont
			{Hussain}},\ and\ \bibinfo {author} {\bibfnamefont {Z.-X.}\ \bibnamefont
			{Shen}},\ }\bibfield  {title} {\bibinfo {title} {Angle-resolved photoemission
			studies of the cuprate superconductors},\ }\href
	{https://doi.org/10.1103/RevModPhys.75.473} {\bibfield  {journal} {\bibinfo
			{journal} {Rev. Mod. Phys.}\ }\textbf {\bibinfo {volume} {75}},\ \bibinfo
		{pages} {473} (\bibinfo {year} {2003})}\BibitemShut {NoStop}%
	\bibitem [{\citenamefont {Proust}\ and\ \citenamefont
		{Taillefer}(2019)}]{CupratesZongshu}%
	\BibitemOpen
	\bibfield  {author} {\bibinfo {author} {\bibfnamefont {C.}~\bibnamefont
			{Proust}}\ and\ \bibinfo {author} {\bibfnamefont {L.}~\bibnamefont
			{Taillefer}},\ }\bibfield  {title} {\bibinfo {title} {The remarkable
			underlying ground states of cuprate superconductors},\ }\href
	{https://doi.org/https://doi.org/10.1146/annurev-conmatphys-031218-013210}
	{\bibfield  {journal} {\bibinfo  {journal} {Annual Review of Condensed Matter
				Physics}\ }\textbf {\bibinfo {volume} {10}},\ \bibinfo {pages} {409}
		(\bibinfo {year} {2019})}\BibitemShut {NoStop}%
	\bibitem [{\citenamefont {Scalapino}(2012)}]{ScalapinoRMP}%
	\BibitemOpen
	\bibfield  {author} {\bibinfo {author} {\bibfnamefont {D.~J.}\ \bibnamefont
			{Scalapino}},\ }\bibfield  {title} {\bibinfo {title} {A common thread: The
			pairing interaction for unconventional superconductors},\ }\href
	{https://doi.org/10.1103/RevModPhys.84.1383} {\bibfield  {journal} {\bibinfo
			{journal} {Rev. Mod. Phys.}\ }\textbf {\bibinfo {volume} {84}},\ \bibinfo
		{pages} {1383} (\bibinfo {year} {2012})}\BibitemShut {NoStop}%
	\bibitem [{\citenamefont {Chen}\ \emph
		{et~al.}(2024{\natexlab{a}})\citenamefont {Chen}, \citenamefont {Choi},
		\citenamefont {Jiang}, \citenamefont {Mei}, \citenamefont {Jiang},
		\citenamefont {Li}, \citenamefont {Agrestini}, \citenamefont
		{Garcia-Fernandez}, \citenamefont {Sun}, \citenamefont {Huang}, \citenamefont
		{Shen}, \citenamefont {Wang}, \citenamefont {Hu}, \citenamefont {Lu},
		\citenamefont {Zhou},\ and\ \citenamefont {Feng}}]{FengRIXS}%
	\BibitemOpen
	\bibfield  {author} {\bibinfo {author} {\bibfnamefont {X.}~\bibnamefont
			{Chen}}, \bibinfo {author} {\bibfnamefont {J.}~\bibnamefont {Choi}}, \bibinfo
		{author} {\bibfnamefont {Z.}~\bibnamefont {Jiang}}, \bibinfo {author}
		{\bibfnamefont {J.}~\bibnamefont {Mei}}, \bibinfo {author} {\bibfnamefont
			{K.}~\bibnamefont {Jiang}}, \bibinfo {author} {\bibfnamefont
			{J.}~\bibnamefont {Li}}, \bibinfo {author} {\bibfnamefont {S.}~\bibnamefont
			{Agrestini}}, \bibinfo {author} {\bibfnamefont {M.}~\bibnamefont
			{Garcia-Fernandez}}, \bibinfo {author} {\bibfnamefont {H.}~\bibnamefont
			{Sun}}, \bibinfo {author} {\bibfnamefont {X.}~\bibnamefont {Huang}}, \bibinfo
		{author} {\bibfnamefont {D.}~\bibnamefont {Shen}}, \bibinfo {author}
		{\bibfnamefont {M.}~\bibnamefont {Wang}}, \bibinfo {author} {\bibfnamefont
			{J.}~\bibnamefont {Hu}}, \bibinfo {author} {\bibfnamefont {Y.}~\bibnamefont
			{Lu}}, \bibinfo {author} {\bibfnamefont {K.-J.}\ \bibnamefont {Zhou}},\ and\
		\bibinfo {author} {\bibfnamefont {D.}~\bibnamefont {Feng}},\ }\bibfield
	{title} {\bibinfo {title} {Electronic and magnetic excitations in la3ni2o7},\
	}\href {https://doi.org/10.1038/s41467-024-53863-5} {\bibfield  {journal}
		{\bibinfo  {journal} {Nature Communications}\ }\textbf {\bibinfo {volume}
			{15}},\ \bibinfo {pages} {9597} (\bibinfo {year}
		{2024}{\natexlab{a}})}\BibitemShut {NoStop}%
	\bibitem [{\citenamefont {Chen}\ \emph
		{et~al.}(2024{\natexlab{b}})\citenamefont {Chen}, \citenamefont {Liu},
		\citenamefont {Jiao}, \citenamefont {Zou}, \citenamefont {Jiang},
		\citenamefont {Li}, \citenamefont {Luo}, \citenamefont {Wu}, \citenamefont
		{Zhang}, \citenamefont {Guo},\ and\ \citenamefont {Shu}}]{uSR}%
	\BibitemOpen
	\bibfield  {author} {\bibinfo {author} {\bibfnamefont {K.}~\bibnamefont
			{Chen}}, \bibinfo {author} {\bibfnamefont {X.}~\bibnamefont {Liu}}, \bibinfo
		{author} {\bibfnamefont {J.}~\bibnamefont {Jiao}}, \bibinfo {author}
		{\bibfnamefont {M.}~\bibnamefont {Zou}}, \bibinfo {author} {\bibfnamefont
			{C.}~\bibnamefont {Jiang}}, \bibinfo {author} {\bibfnamefont
			{X.}~\bibnamefont {Li}}, \bibinfo {author} {\bibfnamefont {Y.}~\bibnamefont
			{Luo}}, \bibinfo {author} {\bibfnamefont {Q.}~\bibnamefont {Wu}}, \bibinfo
		{author} {\bibfnamefont {N.}~\bibnamefont {Zhang}}, \bibinfo {author}
		{\bibfnamefont {Y.}~\bibnamefont {Guo}},\ and\ \bibinfo {author}
		{\bibfnamefont {L.}~\bibnamefont {Shu}},\ }\bibfield  {title} {\bibinfo
		{title} {Evidence of spin density waves in
			${\mathrm{la}}_{3}{\mathrm{ni}}_{2}{\mathrm{o}}_{7\ensuremath{-}\ensuremath{\delta}}$},\
	}\href {https://doi.org/10.1103/PhysRevLett.132.256503} {\bibfield  {journal}
		{\bibinfo  {journal} {Phys. Rev. Lett.}\ }\textbf {\bibinfo {volume} {132}},\
		\bibinfo {pages} {256503} (\bibinfo {year} {2024}{\natexlab{b}})}\BibitemShut
	{NoStop}%
	\bibitem [{\citenamefont {Khasanov}\ \emph {et~al.}(2025)\citenamefont
		{Khasanov}, \citenamefont {Hicken}, \citenamefont {Gawryluk}, \citenamefont
		{Sazgari}, \citenamefont {Plokhikh}, \citenamefont {Sorel}, \citenamefont
		{Bartkowiak}, \citenamefont {B枚tzel}, \citenamefont {Lechermann},
		\citenamefont {Eremin}, \citenamefont {Luetkens},\ and\ \citenamefont
		{Guguchia}}]{uSRsplitDW}%
	\BibitemOpen
	\bibfield  {author} {\bibinfo {author} {\bibfnamefont {R.}~\bibnamefont
			{Khasanov}}, \bibinfo {author} {\bibfnamefont {T.~J.}\ \bibnamefont
			{Hicken}}, \bibinfo {author} {\bibfnamefont {D.~J.}\ \bibnamefont
			{Gawryluk}}, \bibinfo {author} {\bibfnamefont {V.}~\bibnamefont {Sazgari}},
		\bibinfo {author} {\bibfnamefont {I.}~\bibnamefont {Plokhikh}}, \bibinfo
		{author} {\bibfnamefont {L.~P.}\ \bibnamefont {Sorel}}, \bibinfo {author}
		{\bibfnamefont {M.}~\bibnamefont {Bartkowiak}}, \bibinfo {author}
		{\bibfnamefont {S.}~\bibnamefont {B枚tzel}}, \bibinfo {author}
		{\bibfnamefont {F.}~\bibnamefont {Lechermann}}, \bibinfo {author}
		{\bibfnamefont {I.~M.}\ \bibnamefont {Eremin}}, \bibinfo {author}
		{\bibfnamefont {H.}~\bibnamefont {Luetkens}},\ and\ \bibinfo {author}
		{\bibfnamefont {Z.}~\bibnamefont {Guguchia}},\ }\bibfield  {title} {\bibinfo
		{title} {Pressure-enhanced splitting of density wave transitions in
			${\mathrm{la}}_{3}{\text{ni}}_{2}{\mathrm{o}}_{7-\delta}$},\ }\href
	{https://doi.org/10.1038/s41567-024-02754-z} {\bibfield  {journal} {\bibinfo
			{journal} {Nature Physics}\ }\textbf {\bibinfo {volume} {21}},\ \bibinfo
		{pages} {430} (\bibinfo {year} {2025})}\BibitemShut {NoStop}%
	\bibitem [{\citenamefont {Zhao}\ \emph {et~al.}(2025)\citenamefont {Zhao},
		\citenamefont {Zhou}, \citenamefont {Huo}, \citenamefont {Wang},
		\citenamefont {Nie}, \citenamefont {Yang}, \citenamefont {Ying},
		\citenamefont {Wang}, \citenamefont {Wu},\ and\ \citenamefont
		{Chen}}]{ChenxianhuiNMR}%
	\BibitemOpen
	\bibfield  {author} {\bibinfo {author} {\bibfnamefont {D.}~\bibnamefont
			{Zhao}}, \bibinfo {author} {\bibfnamefont {Y.}~\bibnamefont {Zhou}}, \bibinfo
		{author} {\bibfnamefont {M.}~\bibnamefont {Huo}}, \bibinfo {author}
		{\bibfnamefont {Y.}~\bibnamefont {Wang}}, \bibinfo {author} {\bibfnamefont
			{L.}~\bibnamefont {Nie}}, \bibinfo {author} {\bibfnamefont {Y.}~\bibnamefont
			{Yang}}, \bibinfo {author} {\bibfnamefont {J.}~\bibnamefont {Ying}}, \bibinfo
		{author} {\bibfnamefont {M.}~\bibnamefont {Wang}}, \bibinfo {author}
		{\bibfnamefont {T.}~\bibnamefont {Wu}},\ and\ \bibinfo {author}
		{\bibfnamefont {X.}~\bibnamefont {Chen}},\ }\bibfield  {title} {\bibinfo
		{title} {Pressure-enhanced spin-density-wave transition in double-layer
			nickelate ${\mathrm{la}}_{3}{\text{ni}}_{2}{\mathrm{o}}_{7-\delta}$},\ }\href
	{https://www.sciencedirect.com/science/article/pii/S2095927325001811}
	{\bibfield  {journal} {\bibinfo  {journal} {Science Bulletin}\ }\textbf
		{\bibinfo {volume} {70}},\ \bibinfo {pages} {1239} (\bibinfo {year}
		{2025})}\BibitemShut {NoStop}%
	\bibitem [{\citenamefont {Yang}\ \emph
		{et~al.}(2024{\natexlab{a}})\citenamefont {Yang}, \citenamefont {Sun},
		\citenamefont {Hu}, \citenamefont {Xie}, \citenamefont {Miao}, \citenamefont
		{Luo}, \citenamefont {Chen}, \citenamefont {Liang}, \citenamefont {Zhu},
		\citenamefont {Qu}, \citenamefont {Chen}, \citenamefont {Huo}, \citenamefont
		{Huang}, \citenamefont {Zhang}, \citenamefont {Zhang}, \citenamefont {Yang},
		\citenamefont {Wang}, \citenamefont {Peng}, \citenamefont {Mao},
		\citenamefont {Liu}, \citenamefont {Xu}, \citenamefont {Qian}, \citenamefont
		{Yao}, \citenamefont {Wang}, \citenamefont {Zhao},\ and\ \citenamefont
		{Zhou}}]{ZXJNCarpes}%
	\BibitemOpen
	\bibfield  {author} {\bibinfo {author} {\bibfnamefont {J.}~\bibnamefont
			{Yang}}, \bibinfo {author} {\bibfnamefont {H.}~\bibnamefont {Sun}}, \bibinfo
		{author} {\bibfnamefont {X.}~\bibnamefont {Hu}}, \bibinfo {author}
		{\bibfnamefont {Y.}~\bibnamefont {Xie}}, \bibinfo {author} {\bibfnamefont
			{T.}~\bibnamefont {Miao}}, \bibinfo {author} {\bibfnamefont {H.}~\bibnamefont
			{Luo}}, \bibinfo {author} {\bibfnamefont {H.}~\bibnamefont {Chen}}, \bibinfo
		{author} {\bibfnamefont {B.}~\bibnamefont {Liang}}, \bibinfo {author}
		{\bibfnamefont {W.}~\bibnamefont {Zhu}}, \bibinfo {author} {\bibfnamefont
			{G.}~\bibnamefont {Qu}}, \bibinfo {author} {\bibfnamefont {C.-Q.}\
			\bibnamefont {Chen}}, \bibinfo {author} {\bibfnamefont {M.}~\bibnamefont
			{Huo}}, \bibinfo {author} {\bibfnamefont {Y.}~\bibnamefont {Huang}}, \bibinfo
		{author} {\bibfnamefont {S.}~\bibnamefont {Zhang}}, \bibinfo {author}
		{\bibfnamefont {F.}~\bibnamefont {Zhang}}, \bibinfo {author} {\bibfnamefont
			{F.}~\bibnamefont {Yang}}, \bibinfo {author} {\bibfnamefont {Z.}~\bibnamefont
			{Wang}}, \bibinfo {author} {\bibfnamefont {Q.}~\bibnamefont {Peng}}, \bibinfo
		{author} {\bibfnamefont {H.}~\bibnamefont {Mao}}, \bibinfo {author}
		{\bibfnamefont {G.}~\bibnamefont {Liu}}, \bibinfo {author} {\bibfnamefont
			{Z.}~\bibnamefont {Xu}}, \bibinfo {author} {\bibfnamefont {T.}~\bibnamefont
			{Qian}}, \bibinfo {author} {\bibfnamefont {D.-X.}\ \bibnamefont {Yao}},
		\bibinfo {author} {\bibfnamefont {M.}~\bibnamefont {Wang}}, \bibinfo {author}
		{\bibfnamefont {L.}~\bibnamefont {Zhao}},\ and\ \bibinfo {author}
		{\bibfnamefont {X.~J.}\ \bibnamefont {Zhou}},\ }\bibfield  {title} {\bibinfo
		{title} {Orbital-dependent electron correlation in double-layer nickelate
			la3ni2o7},\ }\href {https://doi.org/10.1038/s41467-024-48701-7} {\bibfield
		{journal} {\bibinfo  {journal} {Nature Communications}\ }\textbf {\bibinfo
			{volume} {15}},\ \bibinfo {pages} {4373} (\bibinfo {year}
		{2024}{\natexlab{a}})}\BibitemShut {NoStop}%
	\bibitem [{\citenamefont {Li}\ \emph {et~al.}(2024)\citenamefont {Li},
		\citenamefont {Du}, \citenamefont {Cao}, \citenamefont {Pei}, \citenamefont
		{Zhang}, \citenamefont {Zhai}, \citenamefont {Xu}, \citenamefont {Liu},
		\citenamefont {Li}, \citenamefont {Zhao}, \citenamefont {Li}, \citenamefont
		{Qi}, \citenamefont {Guo}, \citenamefont {Chen},\ and\ \citenamefont
		{Yang}}]{ARPEScpl}%
	\BibitemOpen
	\bibfield  {author} {\bibinfo {author} {\bibfnamefont {Y.}~\bibnamefont
			{Li}}, \bibinfo {author} {\bibfnamefont {X.}~\bibnamefont {Du}}, \bibinfo
		{author} {\bibfnamefont {Y.}~\bibnamefont {Cao}}, \bibinfo {author}
		{\bibfnamefont {C.}~\bibnamefont {Pei}}, \bibinfo {author} {\bibfnamefont
			{M.}~\bibnamefont {Zhang}}, \bibinfo {author} {\bibfnamefont
			{K.}~\bibnamefont {Zhai}}, \bibinfo {author} {\bibfnamefont {R.}~\bibnamefont
			{Xu}}, \bibinfo {author} {\bibfnamefont {Z.}~\bibnamefont {Liu}}, \bibinfo
		{author} {\bibfnamefont {Z.}~\bibnamefont {Li}}, \bibinfo {author}
		{\bibfnamefont {J.}~\bibnamefont {Zhao}}, \bibinfo {author} {\bibfnamefont
			{G.}~\bibnamefont {Li}}, \bibinfo {author} {\bibfnamefont {Y.}~\bibnamefont
			{Qi}}, \bibinfo {author} {\bibfnamefont {H.}~\bibnamefont {Guo}}, \bibinfo
		{author} {\bibfnamefont {Y.}~\bibnamefont {Chen}},\ and\ \bibinfo {author}
		{\bibfnamefont {L.}~\bibnamefont {Yang}},\ }\bibfield  {title} {\bibinfo
		{title} {Electronic correlation and pseudogap-like behavior of
			high-temperature superconductor la$_{3}$ni$_2$o$_{7}$},\ }\href
	{https://doi.org/10.1088/0256-307X/41/8/087402} {\bibfield  {journal}
		{\bibinfo  {journal} {Chin. Phys. Lett.}\ }\textbf {\bibinfo {volume} {41}},\
		\bibinfo {pages} {087402} (\bibinfo {year} {2024})}\BibitemShut {NoStop}%
	\bibitem [{\citenamefont {Shen}\ \emph {et~al.}(2023)\citenamefont {Shen},
		\citenamefont {Qin},\ and\ \citenamefont {Zhang}}]{tJDMRGzhanggm}%
	\BibitemOpen
	\bibfield  {author} {\bibinfo {author} {\bibfnamefont {Y.}~\bibnamefont
			{Shen}}, \bibinfo {author} {\bibfnamefont {M.}~\bibnamefont {Qin}},\ and\
		\bibinfo {author} {\bibfnamefont {G.-M.}\ \bibnamefont {Zhang}},\ }\bibfield
	{title} {\bibinfo {title} {Effective bi-layer model hamiltonian and
			density-matrix renormalization group study for the high-tc superconductivity
			in la3ni2o7 under high pressure},\ }\href
	{https://doi.org/10.1088/0256-307X/40/12/127401} {\bibfield  {journal}
		{\bibinfo  {journal} {Chinese Physics Letters}\ }\textbf {\bibinfo {volume}
			{40}},\ \bibinfo {pages} {127401} (\bibinfo {year} {2023})}\BibitemShut
	{NoStop}%
	\bibitem [{\citenamefont {Jiang}\ \emph {et~al.}(2024)\citenamefont {Jiang},
		\citenamefont {Wang},\ and\ \citenamefont {Zhang}}]{tJKunJiang}%
	\BibitemOpen
	\bibfield  {author} {\bibinfo {author} {\bibfnamefont {K.}~\bibnamefont
			{Jiang}}, \bibinfo {author} {\bibfnamefont {Z.}~\bibnamefont {Wang}},\ and\
		\bibinfo {author} {\bibfnamefont {F.-C.}\ \bibnamefont {Zhang}},\ }\bibfield
	{title} {\bibinfo {title} {High-temperature superconductivity in
			la$_3$ni$_2$o$_7$},\ }\href {https://doi.org/10.1088/0256-307X/41/1/017402}
	{\bibfield  {journal} {\bibinfo  {journal} {Chinese Physics Letters}\
		}\textbf {\bibinfo {volume} {41}},\ \bibinfo {eid} {017402} (\bibinfo {year}
		{2024})}\BibitemShut {NoStop}%
	\bibitem [{\citenamefont {Lange}\ \emph {et~al.}(2024)\citenamefont {Lange},
		\citenamefont {Homeier}, \citenamefont {Demler}, \citenamefont
		{Schollw\"ock}, \citenamefont {Bohrdt},\ and\ \citenamefont
		{Grusdt}}]{tJLange}%
	\BibitemOpen
	\bibfield  {author} {\bibinfo {author} {\bibfnamefont {H.}~\bibnamefont
			{Lange}}, \bibinfo {author} {\bibfnamefont {L.}~\bibnamefont {Homeier}},
		\bibinfo {author} {\bibfnamefont {E.}~\bibnamefont {Demler}}, \bibinfo
		{author} {\bibfnamefont {U.}~\bibnamefont {Schollw\"ock}}, \bibinfo {author}
		{\bibfnamefont {A.}~\bibnamefont {Bohrdt}},\ and\ \bibinfo {author}
		{\bibfnamefont {F.}~\bibnamefont {Grusdt}},\ }\bibfield  {title} {\bibinfo
		{title} {Pairing dome from an emergent feshbach resonance in a strongly
			repulsive bilayer model},\ }\href
	{https://doi.org/10.1103/PhysRevB.110.L081113} {\bibfield  {journal}
		{\bibinfo  {journal} {Phys. Rev. B}\ }\textbf {\bibinfo {volume} {110}},\
		\bibinfo {pages} {L081113} (\bibinfo {year} {2024})}\BibitemShut {NoStop}%
	\bibitem [{\citenamefont {Yang}\ \emph
		{et~al.}(2023{\natexlab{a}})\citenamefont {Yang}, \citenamefont {Zhang},\
		and\ \citenamefont {Zhang}}]{tJYang}%
	\BibitemOpen
	\bibfield  {author} {\bibinfo {author} {\bibfnamefont {Y.-f.}\ \bibnamefont
			{Yang}}, \bibinfo {author} {\bibfnamefont {G.-M.}\ \bibnamefont {Zhang}},\
		and\ \bibinfo {author} {\bibfnamefont {F.-C.}\ \bibnamefont {Zhang}},\
	}\bibfield  {title} {\bibinfo {title} {Interlayer valence bonds and
			two-component theory for high-${T}_{c}$ superconductivity of
			${\mathrm{la}}_{3}{\mathrm{ni}}_{2}{\mathrm{o}}_{7}$ under pressure},\ }\href
	{https://doi.org/10.1103/PhysRevB.108.L201108} {\bibfield  {journal}
		{\bibinfo  {journal} {Phys. Rev. B}\ }\textbf {\bibinfo {volume} {108}},\
		\bibinfo {pages} {L201108} (\bibinfo {year}
		{2023}{\natexlab{a}})}\BibitemShut {NoStop}%
	\bibitem [{\citenamefont {Wang}\ \emph {et~al.}(2025)\citenamefont {Wang},
		\citenamefont {Zhang}, \citenamefont {Jiang},\ and\ \citenamefont
		{Zhang}}]{tJzhangfchun}%
	\BibitemOpen
	\bibfield  {author} {\bibinfo {author} {\bibfnamefont {Z.}~\bibnamefont
			{Wang}}, \bibinfo {author} {\bibfnamefont {H.-J.}\ \bibnamefont {Zhang}},
		\bibinfo {author} {\bibfnamefont {K.}~\bibnamefont {Jiang}},\ and\ \bibinfo
		{author} {\bibfnamefont {F.-C.}\ \bibnamefont {Zhang}},\ }\bibfield  {title}
	{\bibinfo {title} {Self-doped molecular mott insulator for bilayer
			high-temperature superconducting la3ni2o7},\ }\href
	{https://doi.org/10.1093/nsr/nwaf353} {\bibfield  {journal} {\bibinfo
			{journal} {National Science Review}\ }\textbf {\bibinfo {volume} {12}},\
		\bibinfo {pages} {nwaf353} (\bibinfo {year} {2025})}\BibitemShut {NoStop}%
	\bibitem [{\citenamefont {Yang}\ \emph
		{et~al.}(2024{\natexlab{b}})\citenamefont {Yang}, \citenamefont {Oh},\ and\
		\citenamefont {Zhang}}]{tJzhangyh}%
	\BibitemOpen
	\bibfield  {author} {\bibinfo {author} {\bibfnamefont {H.}~\bibnamefont
			{Yang}}, \bibinfo {author} {\bibfnamefont {H.}~\bibnamefont {Oh}},\ and\
		\bibinfo {author} {\bibfnamefont {Y.-H.}\ \bibnamefont {Zhang}},\ }\bibfield
	{title} {\bibinfo {title} {Strong pairing from a small fermi surface beyond
			weak coupling: Application to
			${\mathrm{la}}_{3}{\mathrm{ni}}_{2}{\mathrm{o}}_{7}$},\ }\href
	{https://doi.org/10.1103/PhysRevB.110.104517} {\bibfield  {journal} {\bibinfo
			{journal} {Phys. Rev. B}\ }\textbf {\bibinfo {volume} {110}},\ \bibinfo
		{pages} {104517} (\bibinfo {year} {2024}{\natexlab{b}})}\BibitemShut
	{NoStop}%
	\bibitem [{\citenamefont {Fan}\ \emph {et~al.}(2024)\citenamefont {Fan},
		\citenamefont {Zhang}, \citenamefont {Zhan}, \citenamefont {Lv},
		\citenamefont {Jiang}, \citenamefont {Normand},\ and\ \citenamefont
		{Xiang}}]{tJXiangTaoDwave}%
	\BibitemOpen
	\bibfield  {author} {\bibinfo {author} {\bibfnamefont {Z.}~\bibnamefont
			{Fan}}, \bibinfo {author} {\bibfnamefont {J.-F.}\ \bibnamefont {Zhang}},
		\bibinfo {author} {\bibfnamefont {B.}~\bibnamefont {Zhan}}, \bibinfo {author}
		{\bibfnamefont {D.}~\bibnamefont {Lv}}, \bibinfo {author} {\bibfnamefont
			{X.-Y.}\ \bibnamefont {Jiang}}, \bibinfo {author} {\bibfnamefont
			{B.}~\bibnamefont {Normand}},\ and\ \bibinfo {author} {\bibfnamefont
			{T.}~\bibnamefont {Xiang}},\ }\bibfield  {title} {\bibinfo {title}
		{Superconductivity in nickelate and cuprate superconductors with strong
			bilayer coupling},\ }\href {https://doi.org/10.1103/PhysRevB.110.024514}
	{\bibfield  {journal} {\bibinfo  {journal} {Phys. Rev. B}\ }\textbf {\bibinfo
			{volume} {110}},\ \bibinfo {pages} {024514} (\bibinfo {year}
		{2024})}\BibitemShut {NoStop}%
	\bibitem [{\citenamefont {Lu}\ \emph {et~al.}(2024)\citenamefont {Lu},
		\citenamefont {Pan}, \citenamefont {Yang},\ and\ \citenamefont
		{Wu}}]{LuChenWuCprl}%
	\BibitemOpen
	\bibfield  {author} {\bibinfo {author} {\bibfnamefont {C.}~\bibnamefont
			{Lu}}, \bibinfo {author} {\bibfnamefont {Z.}~\bibnamefont {Pan}}, \bibinfo
		{author} {\bibfnamefont {F.}~\bibnamefont {Yang}},\ and\ \bibinfo {author}
		{\bibfnamefont {C.}~\bibnamefont {Wu}},\ }\bibfield  {title} {\bibinfo
		{title} {Interlayer-coupling-driven high-temperature superconductivity in
			${\mathrm{la}}_{3}{\mathrm{ni}}_{2}{\mathrm{o}}_{7}$ under pressure},\ }\href
	{https://doi.org/10.1103/PhysRevLett.132.146002} {\bibfield  {journal}
		{\bibinfo  {journal} {Phys. Rev. Lett.}\ }\textbf {\bibinfo {volume} {132}},\
		\bibinfo {pages} {146002} (\bibinfo {year} {2024})}\BibitemShut {NoStop}%
	\bibitem [{\citenamefont {Qu}\ \emph {et~al.}(2024)\citenamefont {Qu},
		\citenamefont {Qu}, \citenamefont {Chen}, \citenamefont {Wu}, \citenamefont
		{Yang}, \citenamefont {Li},\ and\ \citenamefont {Su}}]{BilayerSu}%
	\BibitemOpen
	\bibfield  {author} {\bibinfo {author} {\bibfnamefont {X.-Z.}\ \bibnamefont
			{Qu}}, \bibinfo {author} {\bibfnamefont {D.-W.}\ \bibnamefont {Qu}}, \bibinfo
		{author} {\bibfnamefont {J.}~\bibnamefont {Chen}}, \bibinfo {author}
		{\bibfnamefont {C.}~\bibnamefont {Wu}}, \bibinfo {author} {\bibfnamefont
			{F.}~\bibnamefont {Yang}}, \bibinfo {author} {\bibfnamefont {W.}~\bibnamefont
			{Li}},\ and\ \bibinfo {author} {\bibfnamefont {G.}~\bibnamefont {Su}},\
	}\bibfield  {title} {\bibinfo {title} {Bilayer
			${t\text{\ensuremath{-}}J\text{\ensuremath{-}}J}_{\ensuremath{\perp}}$ model
			and magnetically mediated pairing in the pressurized nickelate
			${\mathrm{la}}_{3}{\mathrm{ni}}_{2}{\mathrm{o}}_{7}$},\ }\href
	{https://doi.org/10.1103/PhysRevLett.132.036502} {\bibfield  {journal}
		{\bibinfo  {journal} {Phys. Rev. Lett.}\ }\textbf {\bibinfo {volume} {132}},\
		\bibinfo {pages} {036502} (\bibinfo {year} {2024})}\BibitemShut {NoStop}%
	\bibitem [{\citenamefont {Pan}\ \emph {et~al.}(2024)\citenamefont {Pan},
		\citenamefont {Lu}, \citenamefont {Yang},\ and\ \citenamefont
		{Wu}}]{Wucongcpl2024}%
	\BibitemOpen
	\bibfield  {author} {\bibinfo {author} {\bibfnamefont {Z.}~\bibnamefont
			{Pan}}, \bibinfo {author} {\bibfnamefont {C.}~\bibnamefont {Lu}}, \bibinfo
		{author} {\bibfnamefont {F.}~\bibnamefont {Yang}},\ and\ \bibinfo {author}
		{\bibfnamefont {C.}~\bibnamefont {Wu}},\ }\bibfield  {title} {\bibinfo
		{title} {Effect of rare-earth element substitution in superconducting
			{R$_3$Ni$_2$O$_7$} under pressure},\ }\href
	{https://doi.org/10.1088/0256-307X/41/8/087401} {\bibfield  {journal}
		{\bibinfo  {journal} {Chin. Phys. Lett.}\ }\textbf {\bibinfo {volume} {41}},\
		\bibinfo {pages} {087401} (\bibinfo {year} {2024})}\BibitemShut {NoStop}%
	\bibitem [{\citenamefont {Qin}\ and\ \citenamefont {Yang}(2023)}]{Yangyifeng}%
	\BibitemOpen
	\bibfield  {author} {\bibinfo {author} {\bibfnamefont {Q.}~\bibnamefont
			{Qin}}\ and\ \bibinfo {author} {\bibfnamefont {Y.-f.}\ \bibnamefont {Yang}},\
	}\bibfield  {title} {\bibinfo {title} {High-${T}_{c}$ superconductivity by
			mobilizing local spin singlets and possible route to higher ${T}_{c}$ in
			pressurized ${\mathrm{la}}_{3}{\mathrm{ni}}_{2}{\mathrm{o}}_{7}$},\ }\href
	{https://doi.org/10.1103/PhysRevB.108.L140504} {\bibfield  {journal}
		{\bibinfo  {journal} {Phys. Rev. B}\ }\textbf {\bibinfo {volume} {108}},\
		\bibinfo {pages} {L140504} (\bibinfo {year} {2023})}\BibitemShut {NoStop}%
	\bibitem [{\citenamefont {Luo}\ \emph {et~al.}(2024)\citenamefont {Luo},
		\citenamefont {Lv}, \citenamefont {Wang}, \citenamefont {Wu},\ and\
		\citenamefont {Yao}}]{YaodaotJ}%
	\BibitemOpen
	\bibfield  {author} {\bibinfo {author} {\bibfnamefont {Z.}~\bibnamefont
			{Luo}}, \bibinfo {author} {\bibfnamefont {B.}~\bibnamefont {Lv}}, \bibinfo
		{author} {\bibfnamefont {M.}~\bibnamefont {Wang}}, \bibinfo {author}
		{\bibfnamefont {W.}~\bibnamefont {Wu}},\ and\ \bibinfo {author}
		{\bibfnamefont {D.-X.}\ \bibnamefont {Yao}},\ }\bibfield  {title} {\bibinfo
		{title} {High-tc superconductivity in la3ni2o7 based on the bilayer
			two-orbital t-j model},\ }\href {https://doi.org/10.1038/s41535-024-00668-w}
	{\bibfield  {journal} {\bibinfo  {journal} {npj Quantum Materials}\ }\textbf
		{\bibinfo {volume} {9}},\ \bibinfo {pages} {61} (\bibinfo {year}
		{2024})}\BibitemShut {NoStop}%
	\bibitem [{\citenamefont {Xue}\ and\ \citenamefont
		{Wang}(2024)}]{WangFaCPLpairing}%
	\BibitemOpen
	\bibfield  {author} {\bibinfo {author} {\bibfnamefont {J.-R.}\ \bibnamefont
			{Xue}}\ and\ \bibinfo {author} {\bibfnamefont {F.}~\bibnamefont {Wang}},\
	}\bibfield  {title} {\bibinfo {title} {Magnetism and superconductivity in the
			t閳ユ彄 model of la3ni2o7 under multiband gutzwiller approximation},\
	}\href {https://doi.org/10.1088/0256-307X/41/5/057403} {\bibfield  {journal}
		{\bibinfo  {journal} {Chinese Physics Letters}\ }\textbf {\bibinfo {volume}
			{41}},\ \bibinfo {pages} {057403} (\bibinfo {year} {2024})}\BibitemShut
	{NoStop}%
	\bibitem [{\citenamefont {Sakakibara}\ \emph {et~al.}(2024)\citenamefont
		{Sakakibara}, \citenamefont {Kitamine}, \citenamefont {Ochi},\ and\
		\citenamefont {Kuroki}}]{FLEXsaka}%
	\BibitemOpen
	\bibfield  {author} {\bibinfo {author} {\bibfnamefont {H.}~\bibnamefont
			{Sakakibara}}, \bibinfo {author} {\bibfnamefont {N.}~\bibnamefont
			{Kitamine}}, \bibinfo {author} {\bibfnamefont {M.}~\bibnamefont {Ochi}},\
		and\ \bibinfo {author} {\bibfnamefont {K.}~\bibnamefont {Kuroki}},\
	}\bibfield  {title} {\bibinfo {title} {Possible high ${T}_{c}$
			superconductivity in ${\mathrm{la}}_{3}{\mathrm{ni}}_{2}{\mathrm{o}}_{7}$
			under high pressure through manifestation of a nearly half-filled bilayer
			hubbard model},\ }\href {https://doi.org/10.1103/PhysRevLett.132.106002}
	{\bibfield  {journal} {\bibinfo  {journal} {Phys. Rev. Lett.}\ }\textbf
		{\bibinfo {volume} {132}},\ \bibinfo {pages} {106002} (\bibinfo {year}
		{2024})}\BibitemShut {NoStop}%
	\bibitem [{\citenamefont {Zhang}\ \emph {et~al.}(2023)\citenamefont {Zhang},
		\citenamefont {Lin}, \citenamefont {Moreo}, \citenamefont {Maier},\ and\
		\citenamefont {Dagotto}}]{rpaZhangyangSwave}%
	\BibitemOpen
	\bibfield  {author} {\bibinfo {author} {\bibfnamefont {Y.}~\bibnamefont
			{Zhang}}, \bibinfo {author} {\bibfnamefont {L.-F.}\ \bibnamefont {Lin}},
		\bibinfo {author} {\bibfnamefont {A.}~\bibnamefont {Moreo}}, \bibinfo
		{author} {\bibfnamefont {T.~A.}\ \bibnamefont {Maier}},\ and\ \bibinfo
		{author} {\bibfnamefont {E.}~\bibnamefont {Dagotto}},\ }\bibfield  {title}
	{\bibinfo {title} {Trends in electronic structures and
			${s}_{\ifmmode\pm\else\textpm\fi{}}$-wave pairing for the rare-earth series
			in bilayer nickelate superconductor
			{$\mathrm{R}_{3}\mathrm{Ni}_{2}\mathrm{O}_{7}$}},\ }\href
	{https://doi.org/10.1103/PhysRevB.108.165141} {\bibfield  {journal} {\bibinfo
			{journal} {Phys. Rev. B}\ }\textbf {\bibinfo {volume} {108}},\ \bibinfo
		{pages} {165141} (\bibinfo {year} {2023})}\BibitemShut {NoStop}%
	\bibitem [{\citenamefont {Lechermann}\ \emph {et~al.}(2023)\citenamefont
		{Lechermann}, \citenamefont {Gondolf}, \citenamefont {B\"otzel},\ and\
		\citenamefont {Eremin}}]{EreminRPAandpairing}%
	\BibitemOpen
	\bibfield  {author} {\bibinfo {author} {\bibfnamefont {F.}~\bibnamefont
			{Lechermann}}, \bibinfo {author} {\bibfnamefont {J.}~\bibnamefont {Gondolf}},
		\bibinfo {author} {\bibfnamefont {S.}~\bibnamefont {B\"otzel}},\ and\
		\bibinfo {author} {\bibfnamefont {I.~M.}\ \bibnamefont {Eremin}},\ }\bibfield
	{title} {\bibinfo {title} {Electronic correlations and superconducting
			instability in ${\mathrm{la}}_{3}{\mathrm{ni}}_{2}{\mathrm{o}}_{7}$ under
			high pressure},\ }\href {https://doi.org/10.1103/PhysRevB.108.L201121}
	{\bibfield  {journal} {\bibinfo  {journal} {Phys. Rev. B}\ }\textbf {\bibinfo
			{volume} {108}},\ \bibinfo {pages} {L201121} (\bibinfo {year}
		{2023})}\BibitemShut {NoStop}%
	\bibitem [{\citenamefont {Liu}\ \emph {et~al.}(2023)\citenamefont {Liu},
		\citenamefont {Mei}, \citenamefont {Ye}, \citenamefont {Chen},\ and\
		\citenamefont {Yang}}]{YangfanPRLOdefi}%
	\BibitemOpen
	\bibfield  {author} {\bibinfo {author} {\bibfnamefont {Y.-B.}\ \bibnamefont
			{Liu}}, \bibinfo {author} {\bibfnamefont {J.-W.}\ \bibnamefont {Mei}},
		\bibinfo {author} {\bibfnamefont {F.}~\bibnamefont {Ye}}, \bibinfo {author}
		{\bibfnamefont {W.-Q.}\ \bibnamefont {Chen}},\ and\ \bibinfo {author}
		{\bibfnamefont {F.}~\bibnamefont {Yang}},\ }\bibfield  {title} {\bibinfo
		{title} {${\mathrm{s}}^{\ifmmode\pm\else\textpm\fi{}}$-wave pairing and the
			destructive role of apical-oxygen deficiencies in
			${\mathrm{la}}_{3}{\mathrm{ni}}_{2}{\mathrm{o}}_{7}$ under pressure},\ }\href
	{https://doi.org/10.1103/PhysRevLett.131.236002} {\bibfield  {journal}
		{\bibinfo  {journal} {Phys. Rev. Lett.}\ }\textbf {\bibinfo {volume} {131}},\
		\bibinfo {pages} {236002} (\bibinfo {year} {2023})}\BibitemShut {NoStop}%
	\bibitem [{\citenamefont {Zhang}\ \emph
		{et~al.}(2024{\natexlab{b}})\citenamefont {Zhang}, \citenamefont {Lin},
		\citenamefont {Moreo}, \citenamefont {Maier},\ and\ \citenamefont
		{Dagotto}}]{rpazhangyangNC}%
	\BibitemOpen
	\bibfield  {author} {\bibinfo {author} {\bibfnamefont {Y.}~\bibnamefont
			{Zhang}}, \bibinfo {author} {\bibfnamefont {L.-F.}\ \bibnamefont {Lin}},
		\bibinfo {author} {\bibfnamefont {A.}~\bibnamefont {Moreo}}, \bibinfo
		{author} {\bibfnamefont {T.~A.}\ \bibnamefont {Maier}},\ and\ \bibinfo
		{author} {\bibfnamefont {E.}~\bibnamefont {Dagotto}},\ }\bibfield  {title}
	{\bibinfo {title} {Structural phase transition, s-wave pairing, and magnetic
			stripe order in bilayered superconductor la3ni2o7 under pressure},\ }\href
	{https://doi.org/10.1038/s41467-024-46622-z} {\bibfield  {journal} {\bibinfo
			{journal} {Nature Communications}\ }\textbf {\bibinfo {volume} {15}},\
		\bibinfo {pages} {2470} (\bibinfo {year} {2024}{\natexlab{b}})}\BibitemShut
	{NoStop}%
	\bibitem [{\citenamefont {Braz}\ \emph {et~al.}(2025)\citenamefont {Braz},
		\citenamefont {Martins},\ and\ \citenamefont {da~Silva}}]{rpaBraz}%
	\BibitemOpen
	\bibfield  {author} {\bibinfo {author} {\bibfnamefont {L.~B.}\ \bibnamefont
			{Braz}}, \bibinfo {author} {\bibfnamefont {G.~B.}\ \bibnamefont {Martins}},\
		and\ \bibinfo {author} {\bibfnamefont {L.~G. G. V.~D.}\ \bibnamefont
			{da~Silva}},\ }\bibfield  {title} {\bibinfo {title} {Interlayer interactions
			in ${\mathrm{la}}_{3}{\mathrm{ni}}_{2}{\mathrm{o}}_{7}$ under pressure: From
			${s}^{\ifmmode\pm\else\textpm\fi{}}$ to ${d}_{xy}$-wave superconductivity},\
	}\href {https://doi.org/10.1103/f4wf-56fl} {\bibfield  {journal} {\bibinfo
			{journal} {Phys. Rev. Res.}\ }\textbf {\bibinfo {volume} {7}},\ \bibinfo
		{pages} {033023} (\bibinfo {year} {2025})}\BibitemShut {NoStop}%
	\bibitem [{\citenamefont {Yang}\ \emph
		{et~al.}(2023{\natexlab{b}})\citenamefont {Yang}, \citenamefont {Wang},\ and\
		\citenamefont {Wang}}]{FRGwang}%
	\BibitemOpen
	\bibfield  {author} {\bibinfo {author} {\bibfnamefont {Q.-G.}\ \bibnamefont
			{Yang}}, \bibinfo {author} {\bibfnamefont {D.}~\bibnamefont {Wang}},\ and\
		\bibinfo {author} {\bibfnamefont {Q.-H.}\ \bibnamefont {Wang}},\ }\bibfield
	{title} {\bibinfo {title} {Possible ${s}_{\ifmmode\pm\else\textpm\fi{}}$-wave
			superconductivity in ${\mathrm{la}}_{3}{\mathrm{ni}}_{2}{\mathrm{o}}_{7}$},\
	}\href {https://doi.org/10.1103/PhysRevB.108.L140505} {\bibfield  {journal}
		{\bibinfo  {journal} {Phys. Rev. B}\ }\textbf {\bibinfo {volume} {108}},\
		\bibinfo {pages} {L140505} (\bibinfo {year}
		{2023}{\natexlab{b}})}\BibitemShut {NoStop}%
	\bibitem [{\citenamefont {Gu}\ \emph {et~al.}(2025)\citenamefont {Gu},
		\citenamefont {Le}, \citenamefont {Yang}, \citenamefont {Wu},\ and\
		\citenamefont {Hu}}]{FRGwu}%
	\BibitemOpen
	\bibfield  {author} {\bibinfo {author} {\bibfnamefont {Y.}~\bibnamefont
			{Gu}}, \bibinfo {author} {\bibfnamefont {C.}~\bibnamefont {Le}}, \bibinfo
		{author} {\bibfnamefont {Z.}~\bibnamefont {Yang}}, \bibinfo {author}
		{\bibfnamefont {X.}~\bibnamefont {Wu}},\ and\ \bibinfo {author}
		{\bibfnamefont {J.}~\bibnamefont {Hu}},\ }\bibfield  {title} {\bibinfo
		{title} {Effective model and pairing tendency in the bilayer ni-based
			superconductor ${\mathrm{la}}_{3}{\mathrm{ni}}_{2}{\mathrm{o}}_{7}$},\ }\href
	{https://doi.org/10.1103/PhysRevB.111.174506} {\bibfield  {journal} {\bibinfo
			{journal} {Phys. Rev. B}\ }\textbf {\bibinfo {volume} {111}},\ \bibinfo
		{pages} {174506} (\bibinfo {year} {2025})}\BibitemShut {NoStop}%
	\bibitem [{\citenamefont {Zhan}\ \emph {et~al.}(2026)\citenamefont {Zhan},
		\citenamefont {Le}, \citenamefont {Wu},\ and\ \citenamefont
		{Hu}}]{FRGwu2026}%
	\BibitemOpen
	\bibfield  {author} {\bibinfo {author} {\bibfnamefont {J.}~\bibnamefont
			{Zhan}}, \bibinfo {author} {\bibfnamefont {C.}~\bibnamefont {Le}}, \bibinfo
		{author} {\bibfnamefont {X.}~\bibnamefont {Wu}},\ and\ \bibinfo {author}
		{\bibfnamefont {J.}~\bibnamefont {Hu}},\ }\bibfield  {title} {\bibinfo
		{title} {Impact of nonlocal coulomb repulsion on superconductivity and
			density-wave orders in bilayer nickelates},\ }\bibfield  {journal} {\bibinfo
		{journal} {npj Quantum Materials}\ }\href
	{https://doi.org/10.1038/s41535-026-00883-7} {10.1038/s41535-026-00883-7}
	(\bibinfo {year} {2026})\BibitemShut {NoStop}%
	\bibitem [{\citenamefont {Heier}\ \emph {et~al.}(2024)\citenamefont {Heier},
		\citenamefont {Park},\ and\ \citenamefont {Savrasov}}]{FLEXdvs}%
	\BibitemOpen
	\bibfield  {author} {\bibinfo {author} {\bibfnamefont {G.}~\bibnamefont
			{Heier}}, \bibinfo {author} {\bibfnamefont {K.}~\bibnamefont {Park}},\ and\
		\bibinfo {author} {\bibfnamefont {S.~Y.}\ \bibnamefont {Savrasov}},\
	}\bibfield  {title} {\bibinfo {title} {Competing ${d}_{xy}$ and
			${s}_{\ifmmode\pm\else\textpm\fi{}}$ pairing symmetries in superconducting
			${\mathrm{la}}_{3}{\mathrm{ni}}_{2}{\mathrm{o}}_{7}$:
			$\mathrm{LDA}+\mathrm{FLEX}$ calculations},\ }\href
	{https://doi.org/10.1103/PhysRevB.109.104508} {\bibfield  {journal} {\bibinfo
			{journal} {Phys. Rev. B}\ }\textbf {\bibinfo {volume} {109}},\ \bibinfo
		{pages} {104508} (\bibinfo {year} {2024})}\BibitemShut {NoStop}%
	\bibitem [{\citenamefont {Luo}\ \emph {et~al.}(2023)\citenamefont {Luo},
		\citenamefont {Hu}, \citenamefont {Wang}, \citenamefont {W\'u},\ and\
		\citenamefont {Yao}}]{Yaotbmodel}%
	\BibitemOpen
	\bibfield  {author} {\bibinfo {author} {\bibfnamefont {Z.}~\bibnamefont
			{Luo}}, \bibinfo {author} {\bibfnamefont {X.}~\bibnamefont {Hu}}, \bibinfo
		{author} {\bibfnamefont {M.}~\bibnamefont {Wang}}, \bibinfo {author}
		{\bibfnamefont {W.}~\bibnamefont {W\'u}},\ and\ \bibinfo {author}
		{\bibfnamefont {D.-X.}\ \bibnamefont {Yao}},\ }\bibfield  {title} {\bibinfo
		{title} {Bilayer two-orbital model of
			{$\mathrm{La}_{3}\mathrm{Ni}_{2}\mathrm{O}_{7}$} under pressure},\ }\href
	{https://doi.org/10.1103/PhysRevLett.131.126001} {\bibfield  {journal}
		{\bibinfo  {journal} {Phys. Rev. Lett.}\ }\textbf {\bibinfo {volume} {131}},\
		\bibinfo {pages} {126001} (\bibinfo {year} {2023})}\BibitemShut {NoStop}%
	\bibitem [{\citenamefont {Zhang}\ \emph
		{et~al.}(2024{\natexlab{c}})\citenamefont {Zhang}, \citenamefont {Bai},
		\citenamefont {Kong}, \citenamefont {Wu}, \citenamefont {Xing},\ and\
		\citenamefont {Xu}}]{zhangnjp}%
	\BibitemOpen
	\bibfield  {author} {\bibinfo {author} {\bibfnamefont {H.-Y.}\ \bibnamefont
			{Zhang}}, \bibinfo {author} {\bibfnamefont {Y.-J.}\ \bibnamefont {Bai}},
		\bibinfo {author} {\bibfnamefont {F.-J.}\ \bibnamefont {Kong}}, \bibinfo
		{author} {\bibfnamefont {X.-Q.}\ \bibnamefont {Wu}}, \bibinfo {author}
		{\bibfnamefont {Y.-H.}\ \bibnamefont {Xing}},\ and\ \bibinfo {author}
		{\bibfnamefont {N.}~\bibnamefont {Xu}},\ }\bibfield  {title} {\bibinfo
		{title} {Doping evolution of the normal state magnetic excitations in
			pressurized la3ni2o7},\ }\href {https://doi.org/10.1088/1367-2630/ada0d4}
	{\bibfield  {journal} {\bibinfo  {journal} {New Journal of Physics}\ }\textbf
		{\bibinfo {volume} {26}},\ \bibinfo {pages} {123027} (\bibinfo {year}
		{2024}{\natexlab{c}})}\BibitemShut {NoStop}%
	\bibitem [{\citenamefont {Nie}\ \emph {et~al.}(2026)\citenamefont {Nie},
		\citenamefont {Li}, \citenamefont {Lv}, \citenamefont {Xu}, \citenamefont
		{Jiang}, \citenamefont {Fu}, \citenamefont {Zhou}, \citenamefont {Song},
		\citenamefont {Chen}, \citenamefont {Wang}, \citenamefont {Huang},
		\citenamefont {Lin}, \citenamefont {Jia}, \citenamefont {Shen}, \citenamefont
		{Li}, \citenamefont {Xue},\ and\ \citenamefont {Chen}}]{Nie2026film}%
	\BibitemOpen
	\bibfield  {author} {\bibinfo {author} {\bibfnamefont {Z.}~\bibnamefont
			{Nie}}, \bibinfo {author} {\bibfnamefont {Y.}~\bibnamefont {Li}}, \bibinfo
		{author} {\bibfnamefont {W.}~\bibnamefont {Lv}}, \bibinfo {author}
		{\bibfnamefont {L.}~\bibnamefont {Xu}}, \bibinfo {author} {\bibfnamefont
			{Z.}~\bibnamefont {Jiang}}, \bibinfo {author} {\bibfnamefont
			{P.}~\bibnamefont {Fu}}, \bibinfo {author} {\bibfnamefont {G.}~\bibnamefont
			{Zhou}}, \bibinfo {author} {\bibfnamefont {W.}~\bibnamefont {Song}}, \bibinfo
		{author} {\bibfnamefont {Y.}~\bibnamefont {Chen}}, \bibinfo {author}
		{\bibfnamefont {H.}~\bibnamefont {Wang}}, \bibinfo {author} {\bibfnamefont
			{H.}~\bibnamefont {Huang}}, \bibinfo {author} {\bibfnamefont
			{J.}~\bibnamefont {Lin}}, \bibinfo {author} {\bibfnamefont {J.-F.}\
			\bibnamefont {Jia}}, \bibinfo {author} {\bibfnamefont {D.}~\bibnamefont
			{Shen}}, \bibinfo {author} {\bibfnamefont {P.}~\bibnamefont {Li}}, \bibinfo
		{author} {\bibfnamefont {Q.-K.}\ \bibnamefont {Xue}},\ and\ \bibinfo {author}
		{\bibfnamefont {Z.}~\bibnamefont {Chen}},\ }\bibfield  {title} {\bibinfo
		{title} {Superconductivity and electronic structures of nickelate thin film
			superstructures},\ }\href {https://doi.org/10.1038/s41586-026-10352-7}
	{\bibfield  {journal} {\bibinfo  {journal} {Nature}\ }\textbf {\bibinfo
			{volume} {652}},\ \bibinfo {pages} {628} (\bibinfo {year}
		{2026})}\BibitemShut {NoStop}%
	\bibitem [{\citenamefont {Shen}\ \emph {et~al.}(2026)\citenamefont {Shen},
		\citenamefont {Zhou}, \citenamefont {Miao}, \citenamefont {Li}, \citenamefont
		{Ou}, \citenamefont {Chen}, \citenamefont {Wang}, \citenamefont {Luan},
		\citenamefont {Sun}, \citenamefont {Feng}, \citenamefont {Yong},
		\citenamefont {Li}, \citenamefont {Xu}, \citenamefont {Lv}, \citenamefont
		{Nie}, \citenamefont {Wang}, \citenamefont {Huang}, \citenamefont {Sun},
		\citenamefont {Xue}, \citenamefont {He},\ and\ \citenamefont
		{Chen}}]{Chen2026film}%
	\BibitemOpen
	\bibfield  {author} {\bibinfo {author} {\bibfnamefont {J.}~\bibnamefont
			{Shen}}, \bibinfo {author} {\bibfnamefont {G.}~\bibnamefont {Zhou}}, \bibinfo
		{author} {\bibfnamefont {Y.}~\bibnamefont {Miao}}, \bibinfo {author}
		{\bibfnamefont {P.}~\bibnamefont {Li}}, \bibinfo {author} {\bibfnamefont
			{Z.}~\bibnamefont {Ou}}, \bibinfo {author} {\bibfnamefont {Y.}~\bibnamefont
			{Chen}}, \bibinfo {author} {\bibfnamefont {Z.}~\bibnamefont {Wang}}, \bibinfo
		{author} {\bibfnamefont {R.}~\bibnamefont {Luan}}, \bibinfo {author}
		{\bibfnamefont {H.}~\bibnamefont {Sun}}, \bibinfo {author} {\bibfnamefont
			{Z.}~\bibnamefont {Feng}}, \bibinfo {author} {\bibfnamefont {X.}~\bibnamefont
			{Yong}}, \bibinfo {author} {\bibfnamefont {Y.}~\bibnamefont {Li}}, \bibinfo
		{author} {\bibfnamefont {L.}~\bibnamefont {Xu}}, \bibinfo {author}
		{\bibfnamefont {W.}~\bibnamefont {Lv}}, \bibinfo {author} {\bibfnamefont
			{Z.}~\bibnamefont {Nie}}, \bibinfo {author} {\bibfnamefont {H.}~\bibnamefont
			{Wang}}, \bibinfo {author} {\bibfnamefont {H.}~\bibnamefont {Huang}},
		\bibinfo {author} {\bibfnamefont {Y.-J.}\ \bibnamefont {Sun}}, \bibinfo
		{author} {\bibfnamefont {Q.-K.}\ \bibnamefont {Xue}}, \bibinfo {author}
		{\bibfnamefont {J.}~\bibnamefont {He}},\ and\ \bibinfo {author}
		{\bibfnamefont {Z.}~\bibnamefont {Chen}},\ }\bibfield  {title} {\bibinfo
		{title} {Nodeless superconducting gap and electron-boson coupling in
			{$\mathrm{La,Pr,Sm}_{3}\mathrm{Ni}_{2}\mathrm{O}_{7}$} films},\ }\href
	{https://doi.org/10.1126/science.adw8329} {\bibfield  {journal} {\bibinfo
			{journal} {Science}\ }\textbf {\bibinfo {volume} {392}},\ \bibinfo {pages}
		{1396} (\bibinfo {year} {2026})}\BibitemShut {NoStop}%
	\bibitem [{\citenamefont {Li}\ \emph {et~al.}(2025{\natexlab{b}})\citenamefont
		{Li}, \citenamefont {Zhou}, \citenamefont {Lv}, \citenamefont {Li},
		\citenamefont {Yue}, \citenamefont {Huang}, \citenamefont {Xu}, \citenamefont
		{Shen}, \citenamefont {Miao}, \citenamefont {Song}, \citenamefont {Nie},
		\citenamefont {Chen}, \citenamefont {Wang}, \citenamefont {Chen},
		\citenamefont {Huang}, \citenamefont {Chen}, \citenamefont {Qian},
		\citenamefont {Lin}, \citenamefont {He}, \citenamefont {Sun}, \citenamefont
		{Chen},\ and\ \citenamefont {Xue}}]{XQK2025film}%
	\BibitemOpen
	\bibfield  {author} {\bibinfo {author} {\bibfnamefont {P.}~\bibnamefont
			{Li}}, \bibinfo {author} {\bibfnamefont {G.}~\bibnamefont {Zhou}}, \bibinfo
		{author} {\bibfnamefont {W.}~\bibnamefont {Lv}}, \bibinfo {author}
		{\bibfnamefont {Y.}~\bibnamefont {Li}}, \bibinfo {author} {\bibfnamefont
			{C.}~\bibnamefont {Yue}}, \bibinfo {author} {\bibfnamefont {H.}~\bibnamefont
			{Huang}}, \bibinfo {author} {\bibfnamefont {L.}~\bibnamefont {Xu}}, \bibinfo
		{author} {\bibfnamefont {J.}~\bibnamefont {Shen}}, \bibinfo {author}
		{\bibfnamefont {Y.}~\bibnamefont {Miao}}, \bibinfo {author} {\bibfnamefont
			{W.}~\bibnamefont {Song}}, \bibinfo {author} {\bibfnamefont {Z.}~\bibnamefont
			{Nie}}, \bibinfo {author} {\bibfnamefont {Y.}~\bibnamefont {Chen}}, \bibinfo
		{author} {\bibfnamefont {H.}~\bibnamefont {Wang}}, \bibinfo {author}
		{\bibfnamefont {W.}~\bibnamefont {Chen}}, \bibinfo {author} {\bibfnamefont
			{Y.}~\bibnamefont {Huang}}, \bibinfo {author} {\bibfnamefont {Z.-H.}\
			\bibnamefont {Chen}}, \bibinfo {author} {\bibfnamefont {T.}~\bibnamefont
			{Qian}}, \bibinfo {author} {\bibfnamefont {J.}~\bibnamefont {Lin}}, \bibinfo
		{author} {\bibfnamefont {J.}~\bibnamefont {He}}, \bibinfo {author}
		{\bibfnamefont {Y.-J.}\ \bibnamefont {Sun}}, \bibinfo {author} {\bibfnamefont
			{Z.}~\bibnamefont {Chen}},\ and\ \bibinfo {author} {\bibfnamefont {Q.-K.}\
			\bibnamefont {Xue}},\ }\bibfield  {title} {\bibinfo {title} {Angle-resolved
			photoemission spectroscopy of superconducting (la,pr)3ni2o7/srlaalo4
			heterostructures},\ }\href {https://doi.org/10.1093/nsr/nwaf205} {\bibfield
		{journal} {\bibinfo  {journal} {Natl Sci Rev}\ }\textbf {\bibinfo {volume}
			{12}},\ \bibinfo {pages} {nwaf205} (\bibinfo {year}
		{2025}{\natexlab{b}})}\BibitemShut {NoStop}%
	\bibitem [{\citenamefont {Hao}\ \emph {et~al.}(2025)\citenamefont {Hao},
		\citenamefont {Wang}, \citenamefont {Sun}, \citenamefont {Yang},
		\citenamefont {Mao}, \citenamefont {Yan}, \citenamefont {Sun}, \citenamefont
		{Zhang}, \citenamefont {Han}, \citenamefont {Gu}, \citenamefont {Zhou},
		\citenamefont {Ji},\ and\ \citenamefont {Nie}}]{Hao2025film}%
	\BibitemOpen
	\bibfield  {author} {\bibinfo {author} {\bibfnamefont {B.}~\bibnamefont
			{Hao}}, \bibinfo {author} {\bibfnamefont {M.}~\bibnamefont {Wang}}, \bibinfo
		{author} {\bibfnamefont {W.}~\bibnamefont {Sun}}, \bibinfo {author}
		{\bibfnamefont {Y.}~\bibnamefont {Yang}}, \bibinfo {author} {\bibfnamefont
			{Z.}~\bibnamefont {Mao}}, \bibinfo {author} {\bibfnamefont {S.}~\bibnamefont
			{Yan}}, \bibinfo {author} {\bibfnamefont {H.}~\bibnamefont {Sun}}, \bibinfo
		{author} {\bibfnamefont {H.}~\bibnamefont {Zhang}}, \bibinfo {author}
		{\bibfnamefont {L.}~\bibnamefont {Han}}, \bibinfo {author} {\bibfnamefont
			{Z.}~\bibnamefont {Gu}}, \bibinfo {author} {\bibfnamefont {J.}~\bibnamefont
			{Zhou}}, \bibinfo {author} {\bibfnamefont {D.}~\bibnamefont {Ji}},\ and\
		\bibinfo {author} {\bibfnamefont {Y.}~\bibnamefont {Nie}},\ }\bibfield
	{title} {\bibinfo {title} {Superconductivity in sr-doped la3ni2o7 thin
			films},\ }\href {https://doi.org/10.1038/s41563-025-02327-2} {\bibfield
		{journal} {\bibinfo  {journal} {Nature Materials}\ }\textbf {\bibinfo
			{volume} {24}},\ \bibinfo {pages} {1756} (\bibinfo {year}
		{2025})}\BibitemShut {NoStop}%
	\bibitem [{\citenamefont {Wang}\ \emph {et~al.}(2026)\citenamefont {Wang},
		\citenamefont {Abadi}, \citenamefont {Liu}, \citenamefont {Zhang},
		\citenamefont {Zhong}, \citenamefont {Yu}, \citenamefont {Goodge},
		\citenamefont {Zhang}, \citenamefont {Wu}, \citenamefont {Wang},
		\citenamefont {Li}, \citenamefont {Tarn}, \citenamefont {Ko}, \citenamefont
		{Thampy}, \citenamefont {Lin}, \citenamefont {Hashimoto}, \citenamefont {Lu},
		\citenamefont {Lee}, \citenamefont {Devereaux}, \citenamefont {Jia},
		\citenamefont {Hwang},\ and\ \citenamefont {Shen}}]{Szx2026PRXfilm}%
	\BibitemOpen
	\bibfield  {author} {\bibinfo {author} {\bibfnamefont {B.~Y.}\ \bibnamefont
			{Wang}}, \bibinfo {author} {\bibfnamefont {S.~N.}\ \bibnamefont {Abadi}},
		\bibinfo {author} {\bibfnamefont {Y.}~\bibnamefont {Liu}}, \bibinfo {author}
		{\bibfnamefont {Y.}~\bibnamefont {Zhang}}, \bibinfo {author} {\bibfnamefont
			{Y.}~\bibnamefont {Zhong}}, \bibinfo {author} {\bibfnamefont
			{Y.}~\bibnamefont {Yu}}, \bibinfo {author} {\bibfnamefont {B.~H.}\
			\bibnamefont {Goodge}}, \bibinfo {author} {\bibfnamefont {X.}~\bibnamefont
			{Zhang}}, \bibinfo {author} {\bibfnamefont {Y.-M.}\ \bibnamefont {Wu}},
		\bibinfo {author} {\bibfnamefont {R.}~\bibnamefont {Wang}}, \bibinfo {author}
		{\bibfnamefont {J.}~\bibnamefont {Li}}, \bibinfo {author} {\bibfnamefont
			{Y.}~\bibnamefont {Tarn}}, \bibinfo {author} {\bibfnamefont {E.~K.}\
			\bibnamefont {Ko}}, \bibinfo {author} {\bibfnamefont {V.}~\bibnamefont
			{Thampy}}, \bibinfo {author} {\bibfnamefont {C.}~\bibnamefont {Lin}},
		\bibinfo {author} {\bibfnamefont {M.}~\bibnamefont {Hashimoto}}, \bibinfo
		{author} {\bibfnamefont {D.}~\bibnamefont {Lu}}, \bibinfo {author}
		{\bibfnamefont {Y.~S.}\ \bibnamefont {Lee}}, \bibinfo {author} {\bibfnamefont
			{T.~P.}\ \bibnamefont {Devereaux}}, \bibinfo {author} {\bibfnamefont
			{C.}~\bibnamefont {Jia}}, \bibinfo {author} {\bibfnamefont {H.~Y.}\
			\bibnamefont {Hwang}},\ and\ \bibinfo {author} {\bibfnamefont {Z.-X.}\
			\bibnamefont {Shen}},\ }\bibfield  {title} {\bibinfo {title} {Electronic
			structure of compressively strained bilayer nickelate thin film},\ }\href
	{https://doi.org/10.1103/4h5x-btvx} {\bibfield  {journal} {\bibinfo
			{journal} {Phys. Rev. X}\ }\textbf {\bibinfo {volume} {16}},\ \bibinfo
		{pages} {031008} (\bibinfo {year} {2026})}\BibitemShut {NoStop}%
	\bibitem [{\citenamefont {Sun}\ \emph {et~al.}(2026)\citenamefont {Sun},
		\citenamefont {Jiang}, \citenamefont {Hao}, \citenamefont {Yan},
		\citenamefont {Zhang}, \citenamefont {Wang}, \citenamefont {Yang},
		\citenamefont {Sun}, \citenamefont {Liu}, \citenamefont {Ji}, \citenamefont
		{Gu}, \citenamefont {Zhou}, \citenamefont {Shen}, \citenamefont {Feng},\ and\
		\citenamefont {Nie}}]{NieYF2026NPfilm}%
	\BibitemOpen
	\bibfield  {author} {\bibinfo {author} {\bibfnamefont {W.}~\bibnamefont
			{Sun}}, \bibinfo {author} {\bibfnamefont {Z.}~\bibnamefont {Jiang}}, \bibinfo
		{author} {\bibfnamefont {B.}~\bibnamefont {Hao}}, \bibinfo {author}
		{\bibfnamefont {S.}~\bibnamefont {Yan}}, \bibinfo {author} {\bibfnamefont
			{H.}~\bibnamefont {Zhang}}, \bibinfo {author} {\bibfnamefont
			{M.}~\bibnamefont {Wang}}, \bibinfo {author} {\bibfnamefont {Y.}~\bibnamefont
			{Yang}}, \bibinfo {author} {\bibfnamefont {H.}~\bibnamefont {Sun}}, \bibinfo
		{author} {\bibfnamefont {Z.}~\bibnamefont {Liu}}, \bibinfo {author}
		{\bibfnamefont {D.}~\bibnamefont {Ji}}, \bibinfo {author} {\bibfnamefont
			{Z.}~\bibnamefont {Gu}}, \bibinfo {author} {\bibfnamefont {J.}~\bibnamefont
			{Zhou}}, \bibinfo {author} {\bibfnamefont {D.}~\bibnamefont {Shen}}, \bibinfo
		{author} {\bibfnamefont {D.}~\bibnamefont {Feng}},\ and\ \bibinfo {author}
		{\bibfnamefont {Y.}~\bibnamefont {Nie}},\ }\bibfield  {title} {\bibinfo
		{title} {Observation of superconductivity-induced leading-edge gap in a
			bilayer nickelate},\ }\bibfield  {journal} {\bibinfo  {journal} {Nature
			Physics}\ }\href {https://doi.org/10.1038/s41567-026-03396-z}
	{10.1038/s41567-026-03396-z} (\bibinfo {year} {2026})\BibitemShut {NoStop}%
	\bibitem [{\citenamefont {Christiansson}\ \emph {et~al.}(2023)\citenamefont
		{Christiansson}, \citenamefont {Petocchi},\ and\ \citenamefont
		{Werner}}]{CorrelationJU}%
	\BibitemOpen
	\bibfield  {author} {\bibinfo {author} {\bibfnamefont {V.}~\bibnamefont
			{Christiansson}}, \bibinfo {author} {\bibfnamefont {F.}~\bibnamefont
			{Petocchi}},\ and\ \bibinfo {author} {\bibfnamefont {P.}~\bibnamefont
			{Werner}},\ }\bibfield  {title} {\bibinfo {title} {Correlated electronic
			structure of ${\mathrm{la}}_{3}{\text{ni}}_{2}{\mathrm{o}}_{7}$ under
			pressure},\ }\href {https://doi.org/10.1103/PhysRevLett.131.206501}
	{\bibfield  {journal} {\bibinfo  {journal} {Phys. Rev. Lett.}\ }\textbf
		{\bibinfo {volume} {131}},\ \bibinfo {pages} {206501} (\bibinfo {year}
		{2023})}\BibitemShut {NoStop}%
	\bibitem [{\citenamefont {B\"otzel}\ \emph {et~al.}(2024)\citenamefont
		{B\"otzel}, \citenamefont {Lechermann}, \citenamefont {Gondolf},\ and\
		\citenamefont {Eremin}}]{Botzel}%
	\BibitemOpen
	\bibfield  {author} {\bibinfo {author} {\bibfnamefont {S.}~\bibnamefont
			{B\"otzel}}, \bibinfo {author} {\bibfnamefont {F.}~\bibnamefont
			{Lechermann}}, \bibinfo {author} {\bibfnamefont {J.}~\bibnamefont
			{Gondolf}},\ and\ \bibinfo {author} {\bibfnamefont {I.~M.}\ \bibnamefont
			{Eremin}},\ }\bibfield  {title} {\bibinfo {title} {Theory of magnetic
			excitations in the multilayer nickelate superconductor
			${\mathrm{la}}_{3}{\mathrm{ni}}_{2}{\mathrm{o}}_{7}$},\ }\href
	{https://doi.org/10.1103/PhysRevB.109.L180502} {\bibfield  {journal}
		{\bibinfo  {journal} {Phys. Rev. B}\ }\textbf {\bibinfo {volume} {109}},\
		\bibinfo {pages} {L180502} (\bibinfo {year} {2024})}\BibitemShut {NoStop}%
	\bibitem [{\citenamefont {Lin}\ \emph {et~al.}(2024)\citenamefont {Lin},
		\citenamefont {Zhang}, \citenamefont {Kaushal}, \citenamefont {Alvarez},
		\citenamefont {Maier}, \citenamefont {Moreo},\ and\ \citenamefont
		{Dagotto}}]{rpaLlfDagtoMagPhas}%
	\BibitemOpen
	\bibfield  {author} {\bibinfo {author} {\bibfnamefont {L.-F.}\ \bibnamefont
			{Lin}}, \bibinfo {author} {\bibfnamefont {Y.}~\bibnamefont {Zhang}}, \bibinfo
		{author} {\bibfnamefont {N.}~\bibnamefont {Kaushal}}, \bibinfo {author}
		{\bibfnamefont {G.}~\bibnamefont {Alvarez}}, \bibinfo {author} {\bibfnamefont
			{T.~A.}\ \bibnamefont {Maier}}, \bibinfo {author} {\bibfnamefont
			{A.}~\bibnamefont {Moreo}},\ and\ \bibinfo {author} {\bibfnamefont
			{E.}~\bibnamefont {Dagotto}},\ }\bibfield  {title} {\bibinfo {title}
		{Magnetic phase diagram of a two-orbital model for bilayer nickelates with
			varying doping},\ }\href {https://doi.org/10.1103/PhysRevB.110.195135}
	{\bibfield  {journal} {\bibinfo  {journal} {Phys. Rev. B}\ }\textbf {\bibinfo
			{volume} {110}},\ \bibinfo {pages} {195135} (\bibinfo {year}
		{2024})}\BibitemShut {NoStop}%
	\bibitem [{\citenamefont {Luttinger}(1960)}]{Luttingertheorem}%
	\BibitemOpen
	\bibfield  {author} {\bibinfo {author} {\bibfnamefont {J.~M.}\ \bibnamefont
			{Luttinger}},\ }\bibfield  {title} {\bibinfo {title} {Fermi surface and some
			simple equilibrium properties of a system of interacting fermions},\ }\href
	{https://doi.org/10.1103/PhysRev.119.1153} {\bibfield  {journal} {\bibinfo
			{journal} {Phys. Rev.}\ }\textbf {\bibinfo {volume} {119}},\ \bibinfo {pages}
		{1153} (\bibinfo {year} {1960})}\BibitemShut {NoStop}%
	\bibitem [{\citenamefont {Mazin}\ \emph {et~al.}(2008)\citenamefont {Mazin},
		\citenamefont {Singh}, \citenamefont {Johannes},\ and\ \citenamefont
		{Du}}]{Mazin}%
	\BibitemOpen
	\bibfield  {author} {\bibinfo {author} {\bibfnamefont {I.~I.}\ \bibnamefont
			{Mazin}}, \bibinfo {author} {\bibfnamefont {D.~J.}\ \bibnamefont {Singh}},
		\bibinfo {author} {\bibfnamefont {M.~D.}\ \bibnamefont {Johannes}},\ and\
		\bibinfo {author} {\bibfnamefont {M.~H.}\ \bibnamefont {Du}},\ }\bibfield
	{title} {\bibinfo {title} {Unconventional superconductivity with a sign
			reversal in the order parameter of
			${\mathrm{lafeaso}}_{1\ensuremath{-}x}{\mathrm{f}}_{x}$},\ }\href
	{https://doi.org/10.1103/PhysRevLett.101.057003} {\bibfield  {journal}
		{\bibinfo  {journal} {Phys. Rev. Lett.}\ }\textbf {\bibinfo {volume} {101}},\
		\bibinfo {pages} {057003} (\bibinfo {year} {2008})}\BibitemShut {NoStop}%
	\bibitem [{\citenamefont {Xu}\ and\ \citenamefont {Guterding}(2025)}]{bands31}%
	\BibitemOpen
	\bibfield  {author} {\bibinfo {author} {\bibfnamefont {H.-X.}\ \bibnamefont
			{Xu}}\ and\ \bibinfo {author} {\bibfnamefont {D.}~\bibnamefont {Guterding}},\
	}\bibfield  {title} {\bibinfo {title} {Incommensurate spin fluctuations and
			competing pairing symmetries in
			${\mathrm{la}}_{3}{\mathrm{ni}}_{2}{\mathrm{o}}_{7}$},\ }\href
	{https://doi.org/10.1103/f6nj-34th} {\bibfield  {journal} {\bibinfo
			{journal} {Phys. Rev. B}\ }\textbf {\bibinfo {volume} {112}},\ \bibinfo
		{pages} {174519} (\bibinfo {year} {2025})}\BibitemShut {NoStop}%
	\bibitem [{\citenamefont {Xia}\ \emph {et~al.}(2025)\citenamefont {Xia},
		\citenamefont {Liu}, \citenamefont {Zhou},\ and\ \citenamefont
		{Chen}}]{CrystalNC}%
	\BibitemOpen
	\bibfield  {author} {\bibinfo {author} {\bibfnamefont {C.}~\bibnamefont
			{Xia}}, \bibinfo {author} {\bibfnamefont {H.}~\bibnamefont {Liu}}, \bibinfo
		{author} {\bibfnamefont {S.}~\bibnamefont {Zhou}},\ and\ \bibinfo {author}
		{\bibfnamefont {H.}~\bibnamefont {Chen}},\ }\bibfield  {title} {\bibinfo
		{title} {Sensitive dependence of pairing symmetry on ni-eg crystal field
			splitting in the nickelate superconductor la3ni2o7},\ }\href
	{https://doi.org/10.1038/s41467-025-56206-0} {\bibfield  {journal} {\bibinfo
			{journal} {Nature Communications}\ }\textbf {\bibinfo {volume} {16}},\
		\bibinfo {pages} {1054} (\bibinfo {year} {2025})}\BibitemShut {NoStop}%
	\bibitem [{\citenamefont {Xi}\ \emph {et~al.}(2025)\citenamefont {Xi},
		\citenamefont {Yu},\ and\ \citenamefont {Li}}]{InterlayLi}%
	\BibitemOpen
	\bibfield  {author} {\bibinfo {author} {\bibfnamefont {W.}~\bibnamefont
			{Xi}}, \bibinfo {author} {\bibfnamefont {S.-L.}\ \bibnamefont {Yu}},\ and\
		\bibinfo {author} {\bibfnamefont {J.-X.}\ \bibnamefont {Li}},\ }\bibfield
	{title} {\bibinfo {title} {Transition from
			${s}_{\ifmmode\pm\else\textpm\fi{}}$-wave to
			${d}_{{x}^{2}\ensuremath{-}{y}^{2}}$-wave superconductivity driven by
			interlayer interaction in the bilayer two-orbital model of
			${\text{la}}_{3}{\text{ni}}_{2}{\text{o}}_{7}$},\ }\href
	{https://doi.org/10.1103/PhysRevB.111.104505} {\bibfield  {journal} {\bibinfo
			{journal} {Phys. Rev. B}\ }\textbf {\bibinfo {volume} {111}},\ \bibinfo
		{pages} {104505} (\bibinfo {year} {2025})}\BibitemShut {NoStop}%
\end{thebibliography}
\end{document}